\documentclass[structabstract]{aa}

\usepackage{amsmath}
\usepackage{natbib}
\usepackage{epsfig}
\usepackage{txfonts}
\newcommand{\kms}{\ensuremath{\mathrm{km~s}^{-1}}}

\newcommand{\msun}{\ensuremath{M_\odot}}

\newcommand{\nuc}[2]{\ensuremath{\mathrm{^{#1}#2}}}

\newcommand{\ArdentFLASH}{{\sc Ardent/FLASH}}

\newcommand{\etal}{{et al.\/}\/ }

\bibpunct{(}{)}{;}{a}{}{,}
\begin{document}

\title{Non-spherical core collapse supernovae}
\subtitle{III.~Evolution towards homology and 
          dependence on the numerical resolution}

\author{
        A. Gawryszczak\inst{1} \and
        J. Guzman\inst{2}     \and
        T. Plewa\inst{2}      \and
        K. Kifonidis\inst{3}
       }

\offprints{T. Plewa} 
\mail{tplewa@fsu.edu}

\date{Received / Accepted}      
       
\institute{
  Nicolaus Copernicus Astronomical Center,
  Bartycka 18, 
  00-716 Warsaw, Poland \and
  Department of Scientific Computing,
  Florida State University,
  Tallahassee, FL 32306, U.S.A. \and
  Max-Planck-Institut f\"ur Astrophysik,
  Karl-Schwarzschild-Stra{\ss}e 1, 
  D-85741 Garching, Germany
}

\abstract{} {We study the hydrodynamic evolution of a non-spherical
  core-collapse supernova in two spatial dimensions.  We
  begin our study from the moment of shock revival -- taking into
  account neutrino heating and cooling, nucleosynthesis, convection,
  and the standing accretion shock (SASI) instability of the supernova 
  blast -- and continue for the first week after the explosion when
  the expanding flow becomes homologous and the ejecta enter the 
  early supernova remnant (SNR) phase. We observe the growth and interaction
  of Richtmyer-Meshkov, Rayleigh-Taylor, and Kelvin-Helmholtz
  instabilities resulting in an extensive mixing of the heavy
  elements throughout the ejecta. We obtain a series of
  models at progressively higher resolution and provide a discussion
  of numerical convergence.}
{Different from previous studies, our computations are
  performed in a single domain. Periodic mesh mapping is
  avoided. This is made possible by employing cylindrical
  coordinates, and an adaptive mesh refinement (AMR) strategy in
  which the computational workload (defined as the product of the
  total number of computational cells and the length of the time
  step) is monitored and, if necessary, reduced.}
 {Our results are in overall good agreement with the AMR simulations
  we have reported in the past. We show, however, that
  numerical convergence is difficult to achieve, due to the
  strongly non-linear nature of the problem. Even more importantly,
  we find that our model displays a strong tendency to expand
  laterally away from the equatorial plane and toward the
  poles. We demonstrate that this expansion is a \emph{physical} property
  of the low-mode, SASI instability.  Although the SASI operates
  only within about the first second of the explosion, it leaves
  behind a large lateral velocity gradient in the post shock layer
  which affects the evolution for minutes and hours later. This
  results in a prolate deformation of the ejecta and a fast
  advection of the highest-velocity \nuc{56}{Ni}-rich material from
  moderate latitudes to the polar regions of our grid within only
  300 seconds after core bounce. This effect -- if confirmed by
  3D simulations -- might actually be responsible for the global
  asymmetry of the nickel lines in SN 1987A. Yet, it also poses
  difficulties for the analysis of 2D SASI-dominated explosions in
  terms of the maximum nickel velocities, since discretization
  errors at the poles are considered non-negligible.}
{The simulations demonstrate that significant radial and
  lateral motions in the post-shock region, produced by convective
  overturn and the SASI during the early explosion phase, contribute
  to the evolution for minutes and hours after shock revival. They lead
  to both later clump formation, and a significant prolate
  deformation of the ejecta which are observed even as late as
  one week after the explosion. This ejecta deformation may be
  considered final, since the expansion has long become homologous
  by that time. As pointed out in the recent analysis by Kjaer \etal,
  such an ejecta morphology is in good agreement with the observational
  data of SN 1987A. Systematic future studies are needed to investigate 
  how the SASI-induced late-time lateral expansion that we find in this work
  depends on the dominant mode of the SASI when the early explosion
  phase ends, and to which extent it is affected by the dimensionality of
  the simulations. The impact on and importance of the SASI for
  the distribution of iron group nuclei and 
  the morphology of the young SNR argues for future three-dimensional
  explosion and post-explosion studies on singularity-free grids that
  cover the entire sphere. Given the results of our 2D resolution study,
  present three-dimensional simulations must be regarded as
  underresolved, and their conclusions must be verified by a proper
  numerical convergence analysis in three dimensions.}


\maketitle  

%-------------------------------------------------------------------------
%-------------------------------------------------------------------------
\section{Introduction\label{s:introduction}}
%-------------------------------------------------------------------------
%-------------------------------------------------------------------------

%introduce problem / explain why it is important

More than twenty years after the appearance of SN 1987A, the landmark
event in both modern observational and theoretical supernova (SN)
science, the growing observational SN database continues to challenge
supernova modelers. It is currently accepted that core-collapse
supernovae and their remnants are generally characterized by
large-scale anisotropies, mixing and smaller-scale clumping of
material, together with the penetration of such clumps into the
outermost layers of the ejecta \cite[see e.g.][and references
therein]{hanuschik+88,li+93,aschenbach+95,hughes+00,wang+02,
leonard+06,wang+08,kjaer+10}.

%what others have done 

The general interpretation of these properties of stellar explosions
is that they are the result of hydrodynamic instabilities. Stimulated
mainly by SN 1987A, many research groups have performed simulations of
the late-time (beyond about 1\,s after core bounce) hydrodynamic
evolution of core-collapse SNe in the hope to reproduce and understand
these observations (for a detailed bibliography of early work along
these lines see the first paper of the present series,
\citealt{kifonidis+03}).

Historically, such simulations can be classified into two main
types. Models in which a separate multidimensional neutrino
radiation-hydrodynamic modeling of the supernova explosion mechanism
is attempted, from which the subsequent late-time hydrodynamic
evolution is started, belong to the ``radiation-hydrodynamic explosion
initiation'' or, in short, the RHD approach.  In contrast, models in
which the detailed physics of the ``supernova engine'' is neglected,
and the explosion is artificially triggered by some form of manual
energy deposition into a stellar progenitor model,\footnote{usually in
one spatial dimension by the employment of a piston or a thermal
bomb, and sometimes in two dimensions by the use of manufactured
aspherical initial shock waves} may be assigned to the
``non-radiation-hydrodynamic explosion initiation'' or (in short) the
non-RHD approach.

The non-RHD approach is the older of the two, and has been in
exclusive use -- in many variations -- until the late nineties. It
continues to be used even today in works where the complicated physics
that causes the SN explosion cannot be modeled due to computational or
other constraints, and where the ease in setting up a simulation is of
primary importance, as e.g.\ in performing first studies of
three-dimensional phenomena \citep{hungerford+03,hungerford+05}, in
setting up several different explosions in a number of progenitors
\citep{joggerst+09}, or in the testing of certain hypotheses, such as
the possible formation of jets during the explosion and their impact
on late-time SN evolution \citep{couch+09}.

Unfortunately, the non-RHD approach is unsatisfactory from a
theoretical point of view as it neglects physics which is known to be
important in the supernova context, as e.g.\ neutrino heating and
cooling, and non-radial hydrodynamic instabilities like the standing
accretion shock instability (or SASI,
\citealt{blondin+03,scheck+04,laming07,scheck+08,foglizzo09}), and
multidimensional convection \citep{foglizzo+06,scheck+08}, which occur
within the first second after core bounce. 

This has motivated work on the more advanced models of the RHD type,
which hold the promise to yield a fully consistent hydrodynamic
description of the explosion, once the present computational
difficulties which are connected to neutrino-driven supernova modeling
\citep[cf.][]{Buras+06a,Buras+06b} have been overcome. Such RHD models
were introduced (in two spatial dimensions) by \cite{kifonidis+00} and
improved and extended in the first two installments of the present
series (\citealt{kifonidis+03,kifonidis+06}, henceforth Papers~I and
II, respectively).

At present, both the RHD and the non-RHD approach make use of
parametrizations and assumptions. The level of the employed
approximations is, however, fundamentally different. In the non-RHD
approach the entire \emph{hydrodynamic structure} of a model is
prescribed. A severe consequence of such ``structural
parametrization'', as we will call it in the following, is the
tendency to bias the subsequent evolution with preconceived notions
and assumptions, which are unavoidably introduced in attempts to set
up simple initial data (see the discussion in Papers~I and II). In
contrast, the approximations used in RHD type models mainly center on
the level of accuracy at which the (computationally very expensive)
neutrino physics is treated. The hydrodynamics is then calculated
consistently in response to the resulting neutrino heating. The
approximations made in RHD type models are therefore far less
intrusive than those in models of the non-RHD type.

In Paper~I, for instance, a neutrino-hydrodynamics code which was
based on a simple neutrino light-bulb scheme was used to model the
evolution encompassing the first second of the supernova, given a
boundary condition for the dense, contracting, and neutrino-radiating
neutron star core, which was excised from the computational domain for
reasons of efficiency. 

Paper~II improved upon this treatment, by replacing the neutrino light
bulb with the more accurate, gray neutrino transport scheme developed
by \cite{scheck+06}. Furthermore, the boundary neutrino fluxes were
taken to mimic results of supernova simulations employing full
Boltzmann neutrino transport, but were slightly varied around those
values. This led to a more accurate description of the ratio of
hydrodynamic to heating time scales in the post-shock flow than in the
models of Paper~I, and allowed to properly account for large-scale,
non-radial, low-mode (bi- and quadrupolar) hydrodynamic SASI
instabilities during the first second of the explosion.

In both Papers I and II the late-time hydrodynamics was subsequently
computed with an Adaptive Mesh Refinement (AMR) code. Followed to
about six hours after core bounce, the RHD models presented in Paper~II
have demonstrated great potential to give good agreement with
observations. One particular simulation, model b23a, was successful in
accounting for the most intriguing features of SN 1987A.
Based on a recent analysis of SN 1987A's ejecta morphology
\cite{kjaer+10} in fact argue that only SASI-dominated explosions (such as
model b23a) are fully consistent with the observational data, while
jet-induced explosions, as favored by \cite{wang+02}, are not.

Of course, the additional physics in such RHD models does not come
without cost. Temporal and spatial scales need to be resolved that are
absent in models of the non-RHD type. The computational problems
encountered in trying to account for this physics, along with the need
to evolve the models to times sufficiently late for making comparisons
against observations makes such detailed modeling daunting, already in
two, but even more so in three dimensions.

This is exemplified by the difficulties that \cite{hammer+09}
have experienced in attempting first three-dimensional simulations
of this type. Their runs were initialized from 3D neutrino
radiation-hydrodynamic simulations of the explosion performed by
\cite{scheck07}, which served as data for late-time 3D (purely)
hydrodynamic calculations. Because of limited computational resources,
the 3D neutrino radiation-hydrodynamic models of \cite{scheck07} had
only followed four nuclear species, and lacked a detailed treatment of
nucleosynthesis under non-NSE conditions. Even with this
approximation, his best resolved run could only be evolved to a time
of 0.5 s after bounce, just about half the evolutionary time deemed
necessary in 2D modeling attempts of the explosion phase. The
explosion energy in his model was therefore far from being
saturated. In addition, \cite{scheck07} had considered
boundary neutrino fluxes that would have only given a very
low-energy explosion.

In order to still use this model as initial data for their
subsequent simulations, \cite{hammer+09} artificially changed its
energy distribution to match typical SN explosion
energies of $\sim 1$ Bethe. Unfortunately, the bias for the later
evolution that such manipulation introduces is hard to
quantify. Such modification of an RHD model might, in fact, be viewed
as a regression to structural parametrization.  Moreover, the
resolution that \cite{hammer+09} could afford in three dimensions
using a single-mesh PPM code was naturally substantially inferior to
that which has been achieved in the best-resolved two dimensional
studies that made use of AMR techniques. Nevertheless, their results
are interesting, since they indicate that 2D axisymmetric
simulations overestimate the drag coefficient of newly formed clumps
which are enriched in heavy elements, an effect which has been
predicted for some time (see e.g.\ Paper~I, and \citealt{kane+00}). 
While more accurate and better resolved 3D studies must and will be
performed in the future, there are still a lot of unsettled issues
even in two spatial dimensions.

First, few calculations were carried into the phase of homology, and
in the cases where this has been done
\citep{herant+91,herant+92,herant_woosley94} the late-time models do
not include the early explosion stages in a self-consistent way.
Second, most of the existing models have relatively low resolution and
none of the more recent works known to us has discussed
numerical model convergence, or provided any estimates of accuracy for
such important quantities as the maximum nickel velocities in the
ejecta. As we will show in this paper even highly resolved 2D AMR
models still show variations in the maximum nickel velocities. Third,
in models of the RHD type the effects on these velocities due to
radioactive heating by the decay of nickel and cobalt remain to date
unquantified (in models that were started from simple non-RHD
explosions, \citealt{herant+92} have noticed a mild acceleration due
to radioactive heating). Fourth, parameter studies 
with different stellar models, to investigate the sensitivity
of the results on the density and pressure profiles of the progenitor,
have not yet been performed for models of the RHD type.
And fifth, the effects of fall-back have not
been studied in most simulations, since gravity has either been
completely neglected or treated in the monopole approximation.

%motivation/what we have done 

In the present paper we are going to address the first two of these
points. Our goal is to attain homologous expansion in high-resolution
mesh-based simulations that are initiated from radiation-hydrodynamic
models of the early SN explosion phase, and to provide a resolution
study in order to quantify numerical uncertainties. Ultimately, we
would like to evolve the simulations well into the SN remnant phase,
in order to compare such explosion models -- and hence SN theory --
with a larger observational data base than it has been possible in the
past. In the process we face some new challenges that such late-time
computations pose. In particular we aim to resolve, with limited
computational power, and within a reasonable amount of time, all
crucial physics participating in the evolution.

Our models are based on the $15~\msun$\ blue supergiant star by
\cite{woosley+88}. Within this progenitor is embedded the
SASI-dominated explosion model b23a that was presented in Paper~II. To
allow for computations beyond the shock breakout from the stellar
surface, we have added a stellar wind outside the progenitor star. We
begin our simulations at the final time of the explosion model ($\sim
1$\,s after core bounce), and continue to a final simulation time of 7
days, when the expansion is expected to be completely homologous.

We have structured this paper as follows. In
Sect.~\ref{sect:comp_model} we shall give an overview of our
computational approach, and our initial and boundary conditions. The
results of our simulations will be presented in
Sect.~\ref{sect:results}, starting with a discussion of the dynamics
of spherically symmetric (1-D) models -- which serve as a reference --
and progressing to a high-resolution two-dimensional model.
In Sect.~\ref{sect:discussion} we provide a detailed analysis
of our multidimensional simulations, commenting on aspects such as
the global lateral expansion of our SASI-dominated models, the
resulting anisotropy of the final metal distribution in the ejecta,
the extent of the mixing of heavy elements in mass, the evolution
towards homology, and the numerical convergence in two dimensions.
Section~\ref{sect:summary} finally provides a summary of our findings.

%-------------------------------------------------------------------------
%-------------------------------------------------------------------------
\section{Computational model}
\label{sect:comp_model}
%-------------------------------------------------------------------------
%-------------------------------------------------------------------------

%------------------------
\subsection{Hydrodynamics}
%------------------------

%PPM scheme and Ardent/FLASH 

We are numerically solving Euler's equations of compressible
hydrodynamics in conservative form. We use the Eulerian version of the
Piecewise Parabolic Method (PPM) by \cite{colella+84} as implemented
in \ArdentFLASH, a parallel, block-structured, AMR hydrodynamic code
\citep{fryxell+00}. PPM is a high-order shock-capturing Godunov-type
scheme \citep{godunov59}, is formally second order accurate in space,
and uses Strang-type directional splitting for time advancement
\citep{strang68} to achieve second order accuracy in time.  We are
keeping track of 19 passively advected nuclear species by solving a
separate advection equation for each specie. A list of the species is
given in Table~\ref{t:species}.

\begin{table}
\begin{center}
\caption{Passively advected nuclear species}
\label{t:species}
\begin{tabular}{ccccccc}
\hline

      n        &     p            &   \nuc{1}{H}    &   \nuc{3}{He}    &   \nuc{4}{He}   &   \nuc{12}{C}   &   \nuc{14}{N}  \\
 \nuc{16}{O}   &    \nuc{20}{Ne}  &   \nuc{24}{Mg}  &   \nuc{28}{Si}   &   \nuc{32}{S}   &   \nuc{36}{Ar}  &   \nuc{40}{Ca} \\
 \nuc{44}{Ti}  &    \nuc{48}{Cr}  &   \nuc{52}{Fe}  &   \nuc{54}{Fe}   &   \nuc{56}{Ni}  \\

\hline
\end{tabular}
\end{center}
\end{table}

%Equation of State
The system of  hydrodynamic equations is closed using an electron-positron
equation of state based on table interpolation of the Helmholtz free
energy \citep{timmes+00}.  The Helmholtz equation of state (EoS) takes
into account contributions from electrons and positrons (degenerate
and/or relativistic), radiation pressure, and ions assuming complete
ionization. The physical limits of this EoS are $10^{-10} < \rho <
10^{11}$ g~cm$^{-3}$, and $10^4 < T < 10^{11}$~K, appropriate for
stellar interiors and the early stages of post-explosion supernova
remnant evolution. To allow for the evolution at very late times, we have
extended the original Helmholtz EoS to the low-density, low-temperature
regime, by blending it smoothly with an ideal gas equation of state.

%------------------------
\subsection{Simulation setup}
%------------------------

%Spherical grid vs Cylindrical
Previous simulations of multidimensional core-collapse supernovae have
been typically performed in spherical coordinates, which severely
limits the time step due to the small grid spacing towards the center,
combined with the extreme flow conditions prevailing in that region.
Another drawback of using spherical geometry is the fact that while
one may be resolving the interior of the star very well, the
resolution is quickly lost (especially in the angular direction) at large
radii. Thus the resolution is also highly nonuniform, making
convergence analysis difficult.  Also, while spherical geometry might
be optimal at early times of the simulation, it is not desirable at
later times when the dynamics we are most interested in is taking
place far away from the center. Yet another disadvantage of using
spherical geometry is the presence of a grid singularity at the center
of the grid, which in most cases is avoided by using a
computational domain whose inner radial boundary is located at a
finite radius. We avoid most problems associated with a spherical
mesh by using a cylindrical grid. Also, a cylindrical mesh can be easily 
combined with adaptive mesh refinement, avoiding cumbersome
remeshing/restart procedures, and allowing us to take the simulations
farther in time than we otherwise would have been able to.

%Boundary Conditions, Points mass, and gravitational potential
Our two-dimensional cylindrical grid covers the region $[0, 1.01\times
10^{15}]$~cm in radius and $[-1.01\times 10^{15}, 1.01\times
10^{15}]$~cm in the vertical direction. We impose a reflecting boundary
condition at $R=0$ (symmetry axis) and allow for free flow everywhere
else. The center of the star, where the neutron star is forming, is
approximated by a non-moving gravitating sink hole. The hole is
treated as a point-mass located at the grid center with an initial
mass of $1.099~\msun$. It occupies a semi-circular region extending up
to 3 computational cells from the center. The velocity of the material
inside the hole is reset to zero after every step.  The density inside
the sink hole is kept constant ($\approx 10^{-3}$ of the density of
the material just outside the hole) and the mass of material accreted
into the hole is added to the central point mass (representing the
neutron star). The gravity of the central mass and the self-gravity of
the mass residing on the grid are calculated in the spherically 
symmetric (monopole) approximation.

%Initial Conditions: explosion component, progenitor, and stellar wind
At the start of our simulation the grid is populated with three
distinct components. Excluding the inner core occupied by the sink
hole, the {\it explosion component} extends up to a radius of
$r=1.7\times10^9$~cm. (In what follows, the radial distance from the
grid origin is denoted as $r$.) The explosion component is the b23a
model at a time of $t=0.92$~s after core bounce, that was obtained in
Paper~II. Model b23a in turn is based on the WPE15 ls (180)
post-collapse model of the $15~\msun$\ progenitor of \cite{woosley+88}
presented by \cite{bruenn93}. 

In an attempt to minimize a possible contribution and
accumulation of numerical discretization errors near the polar axis
of the explosion model, two 10-degree wide sections near the axis of
model b23a were overwritten by mirror-reflections of the abutting
mesh regions (i.e.\ in these regions the density, for instance, was
initialized according to $\rho(\theta)=\rho(20^\circ-\theta)$).
However, pronounced axial flows still developed during the
subsequent evolution. As we will show below, these axial flows are
promoted by the global lateral expansion of that SASI explosion
model. We recognize that the flow dynamics near the symmetry axis
is due to a blend of a real physical effect, the fact that the
degrees of freedom of the flow are restricted, and the likely
presence of non-negligible discretization errors in these regions (see
Sect.~\ref{sect:discussion}).

The {\it progenitor model} is identical to that which we already used in
Paper~II (S.~E.~Woosley, private communication). The model envelope
extends up to a radius of $\approx 3.9\times 10^{12}$~cm. The
remaining portion of the computational domain is filled with a {\it
  stellar wind}. The wind is spherically symmetric with a velocity of
$15$~\kms, a temperature of $1\times 10^4$ K, and a density
corresponding to a constant mass loss rate of $1\times
10^{-5}~\msun$\ yr$^{-1}$. To map the explosion and progenitor
component to our computational domain we employ monotonic cubic
interpolation \citep{steffen90}.

%------------------------
\subsection{Computational strategy}
%------------------------

%refinement criteria, refinement strategy
Since \ArdentFLASH\ uses AMR we are able to study the hydrodynamic
evolution with many levels of refinement. In our case the mesh is refined
if there are density jumps $> 0.2$, pressure jumps $> 0.4$, or if the
abundance of nickel exceeds 10\% by mass. We initially start the
simulations with a large number of refinement levels and then remove
these levels depending on the instantaneous number of zones (to limit
memory use), and the calculated step size (to limit computational
time). The latter two refinement criteria allow us to control the
total computational cost of an individual model.

\begin{figure}[ht!]
  \begin{center}
    \includegraphics[bb=91 226 512 638,width=0.45\textwidth,clip=true]{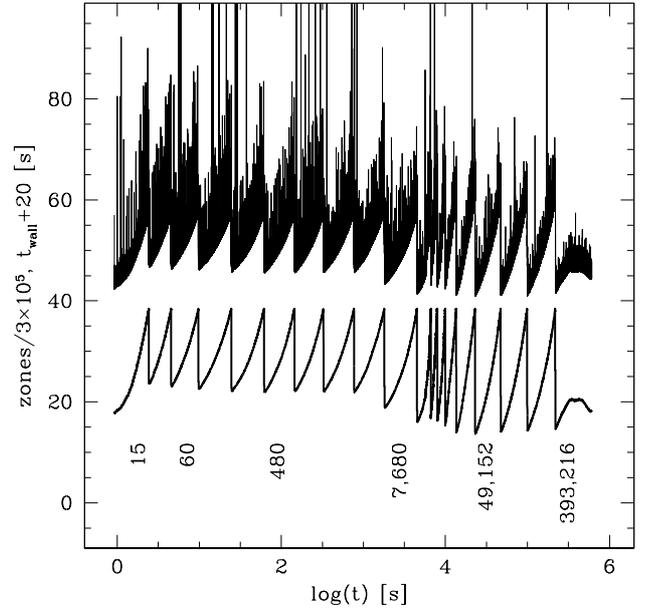}
    \caption{Computational workload in the two-dimensional 15~km
      resolution model of a supernova explosion as a function of the
      logarithm of the simulation time.  The number of zones is shown
      with a thick solid line (bottom curve).  The simulation begins
      with an effective resolution of $15$~km which is gradually
      decreased with time; at the final time the effective model
      resolution is $393,216$~km. The wall-clock time per double
      time-step is shown with a thin line (upper curve). Rapid
      variations in the wall-clock time are partly due to mesh refinement
      ($\approx15$\% amplitude change) and code I/O (large amplitude
      spikes). The simulation has been performed at DOE's NERSC on the
      Franklin Cray XT4 supercomputer using 32 cores.}
      \label{f:refinement_2d_15km_e}
  \end{center}
\end{figure}

%talk more about refinement vs time
In Fig.~\ref{f:refinement_2d_15km_e}
%
%
%
%%\begin{comment}
%%\end{comment}
%
%
%
we show the number of mesh cells (bottom curve) as a function of the
simulation time.  The top curve in this figure shows the wall-clock
time for every double time-step in the simulation. We note $4-5$
second variations ($\approx 15$\% amplitude) in the wall-clock time
due to mesh adaptation, which takes place every other double time
step. Erratic variations of much larger ($\approx 20-30$~s) amplitude
correspond to writing output files (extreme cases are due to
I/O congestion on the computing system).

\begin{figure*}[ht!]
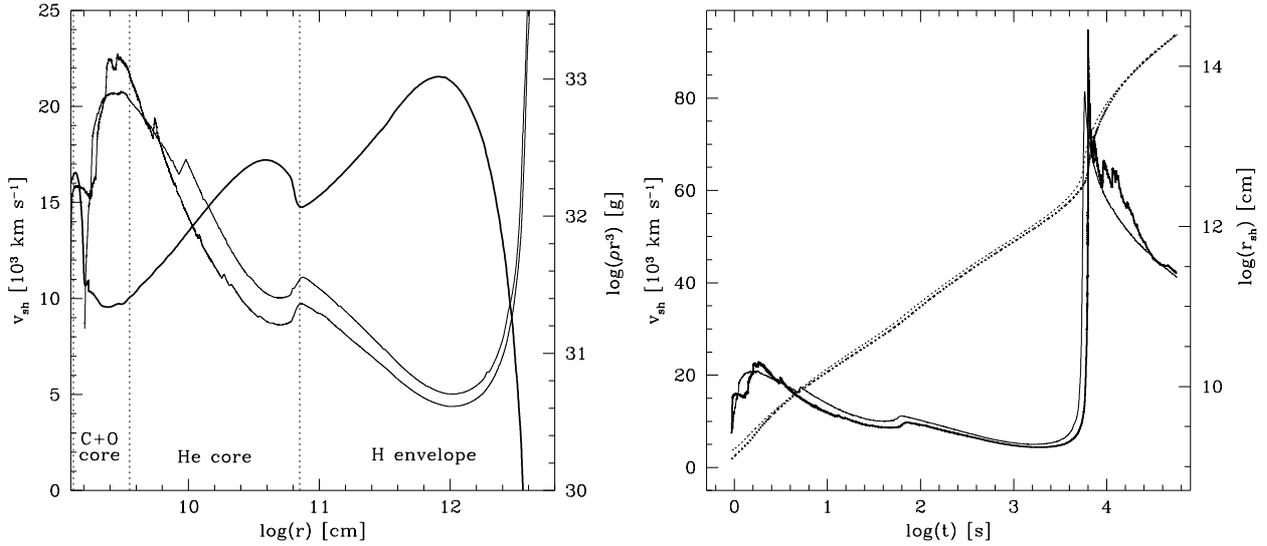

  \begin{center}
    \includegraphics[bb=88 224 566 644,width=0.45\textwidth,clip=true]{13431fg2.epsi}
    \includegraphics[bb=88 224 566 644,width=0.45\textwidth,clip=true]{13431fg3.epsi}
    \caption{Supernova shock dynamics in one- and two-dimensional
      explosion models. (left panel) Evolution of the shock speed
      (left scale) as a function of radius in the one-dimensional
      (thin) and two-dimensional (medium line) $15$~km model. The run of
      the mass distribution $(\rho r^3)$ (right scale) in the
      progenitor star is shown with a thick solid line. Note the
      tight correlation between the shock speed and the mass
      distribution. (right panel) Temporal evolution of the supernova
      shock speed (solid lines) and radius (dotted lines) in the
      one-dimensional (thin lines) and two-dimensional (thick lines)
      $15$~km models. Note the rapid shock acceleration after the
      shock reaches the progenitor's outermost layers around
      $\log(t)\approx 3.75$.}
    \label{f:shockSpeed}
  \end{center}
\end{figure*}

The general semi-periodic shape of the curves reflects our refinement
strategy. During each cycle occupying several hundred time steps, the
number of computational cells increases. Once the maximum allowed
number of cells is reached, the finest level of refinement is
automatically removed and the computation proceeds without
interruption. The mesh level removal is done gradually over several
time steps to avoid problems with mapping the mesh hierarchy to the
process space. Occasionally, the flow conditions may change such that
the time step is reduced below a certain threshold (e.g.\ during shock
acceleration at $\log(t)\approx 3.75$, see also
Fig.~\ref{f:shockSpeed}
%
%
%
%\begin{comment}
%\end{comment}
%
%
%
below) and significantly many more mesh cells are removed to limit the
computational workload. The total computational cost of the 15~km
model is $\approx 1.02\times 10^5$ steps and $\approx 172$ hours on
NERSC's Franklin Cray XT4 with 32 cores.

%-------------------------------------------------------------------------
%-------------------------------------------------------------------------
\section{Model results}
\label{sect:results}
%-------------------------------------------------------------------------
%-------------------------------------------------------------------------

Although the main focus of this work is the effects of two-dimensional
hydrodynamic instabilities, we start with one-dimensional models as
these are useful for understanding the gross overall dynamics of the
system, serve as a reference point in the development of
two-dimensional models, and -- very importantly in the present context
-- help optimizing computational strategies in adaptive simulations.
In two spatial dimensions fluid instabilities typically grow at
contact discontinuities due to the preexisting flow perturbations and
the interaction with the passing shock wave. This complicates the flow
pattern, and usually renders the models too costly for making numerous
adjustments to the simulation strategy.

We expect that our simulations will be largely consistent with those
reported in Paper~II. This is because several essential components of
the two models, both physics-related and computational, are
common. However, we anticipate to observe also some differences due to
the highly non-linear nature of the problem, the use of a different
grid geometry, the lack of mesh remapping, and the overall
higher mesh resolution.

%------------------------
\subsection{Spherically symmetric model\label{ss:1d}}
%------------------------

\begin{figure}[ht!]
  \begin{center}
    \includegraphics[bb=48 224 527 640,width=0.45\textwidth,clip=true]{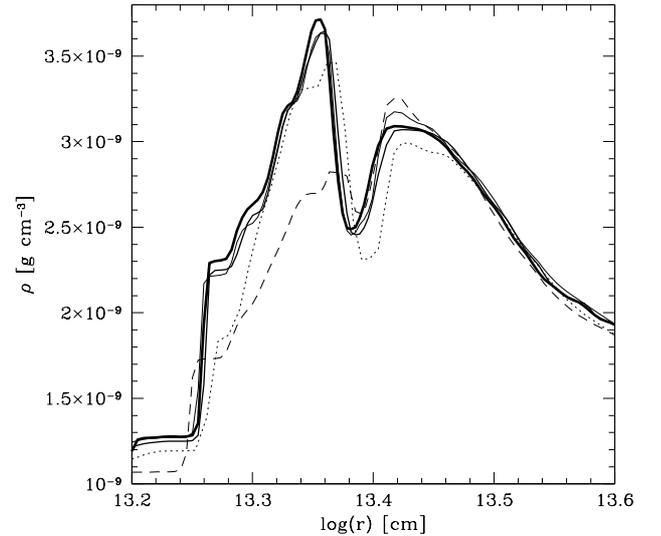}
    \caption{Density profiles in log scale as a function of the
      logarithm of the radius in the densest part of the ejecta in
      one-dimensional models at the final time ($t=7$ days). Results
      are shown for models with a resolution of $15$ (thick solid),
      $30$ (medium solid), $60$ (thin solid), $120$ (dotted), and
      $240$~km (dashed line). Note the gradual decrease in the
      density variation as the mesh resolution increases, which is
      indicative of numerical convergence. The density profiles of models with
      a resolution of at least $60$~km appear very closely matched.}
    \label{f:dens_7days_1d}
  \end{center}
\end{figure}

We have computed one-dimensional models at effective resolutions of
240, 120, 60, 30, and 15~km. Figure~\ref{f:dens_7days_1d}
%
%
%
%\begin{comment}
%\end{comment}
%
%
%
shows the density distribution as a function of the logarithm of the
radius in these one-dimensional models at $t=7$ days.  As we increase
the effective resolution from 240~km (dashed line in
Fig.~\ref{f:dens_7days_1d}) to 120~km (dotted line), the density
increases by $\approx 25$\% at the first density maximum ($\log(r)
\approx 13.35$) and decreases by $\approx 10$\% at the second maximum
($\log(r) \approx 13.42$). Doubling the resolution to $60$~km (thin
solid line) results in a mild density increase at both maxima by
$<10$\%. The density at the second maximum oscillates with the
resolution with a progressively smaller amplitude, while more
monotonic convergence is observed at the first maximum. Despite the
aforementioned variations, the one-dimensional density profile seems
fairly well established already at a resolution of $60$~km.

\begin{figure}[ht!]
  \begin{center}
    \includegraphics[bb=88 224 514 640,width=0.45\textwidth,clip=true]{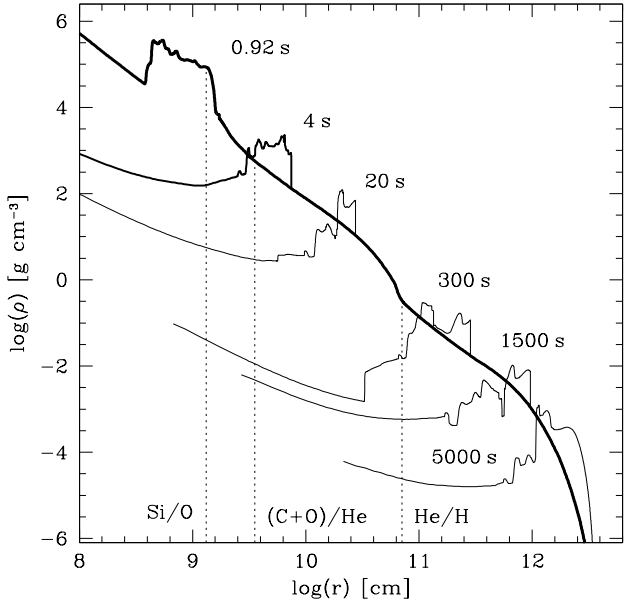}
    \caption{Evolution of the density in log scale before shock
      breakout for the one-dimensional explosion model at an effective
      resolution of $15$~km. The initial positions of the Si/O,
      (C+O)/He, and He/H composition interfaces are marked by dotted
      vertical lines.}
    \label{f:logd_1d_15km}
  \end{center}
\end{figure}

%describe plot of density in log-log scale (up to 5e3 seconds) [K+03.F8a, JG]\\
We begin our simulation at $t=0.92$~s after core bounce with the
supernova shock located just past the Si/O interface at $\log(r)\approx
9.2$~cm (see Fig.~\ref{f:logd_1d_15km}).
%
%
%
%\begin{comment}
%\end{comment}
%
%
%
Behind the supernova shock one can notice a reverse shock at
$\log(r)\approx 8.6$~cm that was formed when the main shock entered
the oxygen core (cf. Paper~I). In between the two shocks sits a
dense shell of metal-rich material which contains most of the ejecta
mass. The dense shell is moving outward due in part to its momentum,
and the combined action of pressure gradients in the shocked gas, the
gravity of the central mass, and the self-gravity of the
progenitor. The reverse shock present at $t=0.92$~s eventually
disappears from the grid as it passes through the inner region (sink
hole), but is partially reflected. In addition to the original reverse
shock, two more reverse shocks form during the passage of the
supernova shock through the envelope. The first of those
reverse shocks originates when the supernova shock enters the
helium shell, and the second one after the shock enters the
hydrogen envelope. Those shocks can be seen in
Fig.~\ref{f:logd_1d_15km} at $t=20$~s and $t=1500$~s at radius
$\log(r)\approx 10.25$~cm and $\log(r)\approx 11.68$~cm,
respectively. They form when the leading shock first
accelerates and then decelerates, which means the freshly shocked
material is moving slower than the shocked material at earlier
time. This leads to piling up of the shocked material, and local
density and pressure increase. If the conditions are right, and the
shocked material moves supersonically with respect to the dense
region, this acoustic perturbation becomes a reverse shock
\citep{herant+94}. Similarly to the original reverse shock, the
additional reverse shocks are reflected off the central region.

%describe evolution of shock speed [K+03.F8b, TP]\\
A shock wave accelerates (decelerates) when it travels through a
medium with the density decreasing faster (slower) than $\propto
r^{-3}$ \citep{sedov59,herant+94}. In other words the shock speed
depends on the mass overrun by the shock, $v_\mathrm{sh}\propto(\rho
r^3)$. Due to the density structure of the progenitor, the supernova
shock propagates through the star in an unsteady fashion. It is the
unsteady propagation of the supernova shock that is responsible for
Rayleigh-Taylor instabilities that develop along material interfaces
\citep{chevalier+78,herant+94}. The propagation of the blast through
the star is illustrated in Fig.~\ref{f:shockSpeed}. While moving
through the C+O core, the shock initially accelerates to slightly over
$20,000$~\kms. Then it slows down in its motion through the helium
shell to $\approx 10,000$~\kms. Upon entering the hydrogen envelope,
it briefly speeds up, and subsequently decelerates again, to $\approx
5500$~\kms. Around $\log(t)\approx 3.2$ ($t\approx 1,600$~s) the shock
begins to accelerate in the outer stellar layers, ultimately reaching
velocities of $\approx 100,000$~\kms. Once the blast has left the
envelope, it gradually slows down inside the stellar wind. This last
acceleration/deceleration sequence gives rise to the formation of a
final reverse shock, and is responsible for the growth of a
Rayleigh-Taylor instability at the interface between the shocked
ejecta and the shocked wind (see below). The shock is traveling at
$\approx 32,500$~\kms\ when it leaves the computational domain at
$t\approx 3$ days.

%------------------------
\subsection{Two-dimensional model\label{ss:2d}}
%------------------------

%describe density morphology plots [TP] (4, 20, 300, 1500, 5000, 10000, 1 d, 7 d)
Although one-dimensional models provide invaluable information about
several major characteristics of the exploding star, qualitatively new
phenomena emerge in multidimensions. These are chiefly related to
fluid flow instabilities developing at the material interfaces that
separate various nuclear species. In what follows, our intention
is to highlight the richness and complexity of the mixing process
rather than to provide a quantitative comparison of contributions of
the individual participating hydrodynamic instabilities. We feel the
value of such a \emph{quantitative} comparison would be extremely
limited and could possibly produce confusion given the long list of
uncertainties involved in the physical model (progenitor) or numerics
(mesh resolution, assumed symmetry). We therefore think it is justified
to offer solely a qualitative discussion.

The main process leading to mixing of different elements 
in supernovae is the Rayleigh-Taylor instability
\cite[RTI;][]{chandrasekhar61,sharp84,youngs84,sadot+05}. The RTI requires
the presence of a material interface separating fluids at two
different densities and a sustained acceleration pointing across the
interface from lighter to denser material. This situation naturally
arises in thermonuclear supernovae, where hot and light ashes of
nuclear fuel buoyantly expand into unburned material
\citep{nomoto+76,mueller+86,khokhlov95}. In core-collapse supernovae,
however, the acceleration at the material interfaces is due to a
positive pressure gradient that results from the acceleration and
deceleration phases of the supernova shock's motion. As we discussed
in the previous section, the supernova shock experiences two major
deceleration episodes during its propagation inside the envelope, and
one more deceleration phase after it leaves the star and sweeps
through the wind (see right panel in Fig.~\ref{f:shockSpeed}). Each
deceleration phase creates a positive pressure gradient in layers with
a negative density gradient, i.e.\ conditions suitable for RTI.

Besides the RTI, two more fluid instabilities are at work in our case.
The Richtmyer-Meshkov instability (RMI; \citealt{richtmyer60,meshkov69,
brouillette02}) occurs at
material interfaces subjected to an impulsive acceleration that (in the
core-collapse supernova setting) is provided by the supernova shock
and also a series of reverse shocks sweeping through the
ejecta. Although the overall radial expansion clearly dominates the
explosion dynamics, initial flow nonuniformities and later
contributions from the RTI and RMI produce velocity shear, naturally
inducing the Kelvin-Helmholtz instability
\citep[KHI;][]{lamb32,chandrasekhar61,sohn+05,youngs+08}. Both the RTI
and RMI, as well as the KHI, grow from small-scale features in the
current problem and thus their role can be quantified only in
well-resolved models. To our knowledge, no such quantitative
comparison of contributions of various instabilities has ever been
done. It is also beyond the scope of the present study.

In our two-dimensional models we can only observe the combined action
of all three instabilities. An assessment of the relative
contributions of the RMI and the RTI based solely on the flow
morphology is an extremely daunting task. At the end of the linear
phase, both instabilities produce very similar structures differing
only in details \cite[see e.g.][]{abarzhi+03}. We may expect, however, that different
instabilities will dominate the evolution at various times. 
The RMI will contribute only after a sufficiently aspherical shock
interacts with a (radial) density gradient. The RTI contribution will
be non-vanishing only in zones with pressure and density gradients of
opposite signs. If such conditions can be sustained for a sufficiently
long time, the RTI with its asymptotic $t^2$ growth will (locally)
outpace the RMI contribution (which scales as $\approx t^{0.3}$). Yet,
the present simulations -- and the RTI growth rates presented in
Paper~I -- indicate that this is not universally the case.
In particular at the He/H interface of our progenitor model,
RMI-growth dominates.
This can be understood by considering that the initial growth time
scales are, $\tau_{RTI} \sim \text{(length scale/effective
acceleration)}^\frac{1}{2}$, and $\tau_{RMI} \sim \text{(length
scale)/(velocity difference)}$. The velocity difference (postshock vs
unshocked envelope) is by far greater than the (effective rate of)
deceleration at the H/He interface. Also, small scale perturbations
grow faster in the RMI than in the RTI. This latter factor is further
amplified by the RMI which initially decreases the amplitude of the
preexisting perturbations in case of He/H (heavy/light) interfaces
\cite[see][Sect. 3.1]{brouillette02}. Both factors result in the RMI
dominating over the RTI, at least initially.
It can also be expected that due to the relatively weak shear along
the interfaces, the KHI growth will be modest, will occur at late
times, and have less impact than either the RTI or RMI on the
evolution.

In our subsequent discussion of the evolution in multidimensions we
will focus on the best resolved model obtained at a resolution of
$15$~km. We will defer commenting on numerical convergence to
Sect.~\ref{ss:convergence} below. Figure~\ref{f:snm_r15be_early_cl}
%
%
%
%%\begin{comment}
\begin{figure}[ht!]
  \begin{center}
    \includegraphics[bb=75 571 1154 1099,width=0.45\textwidth,clip=true]
    {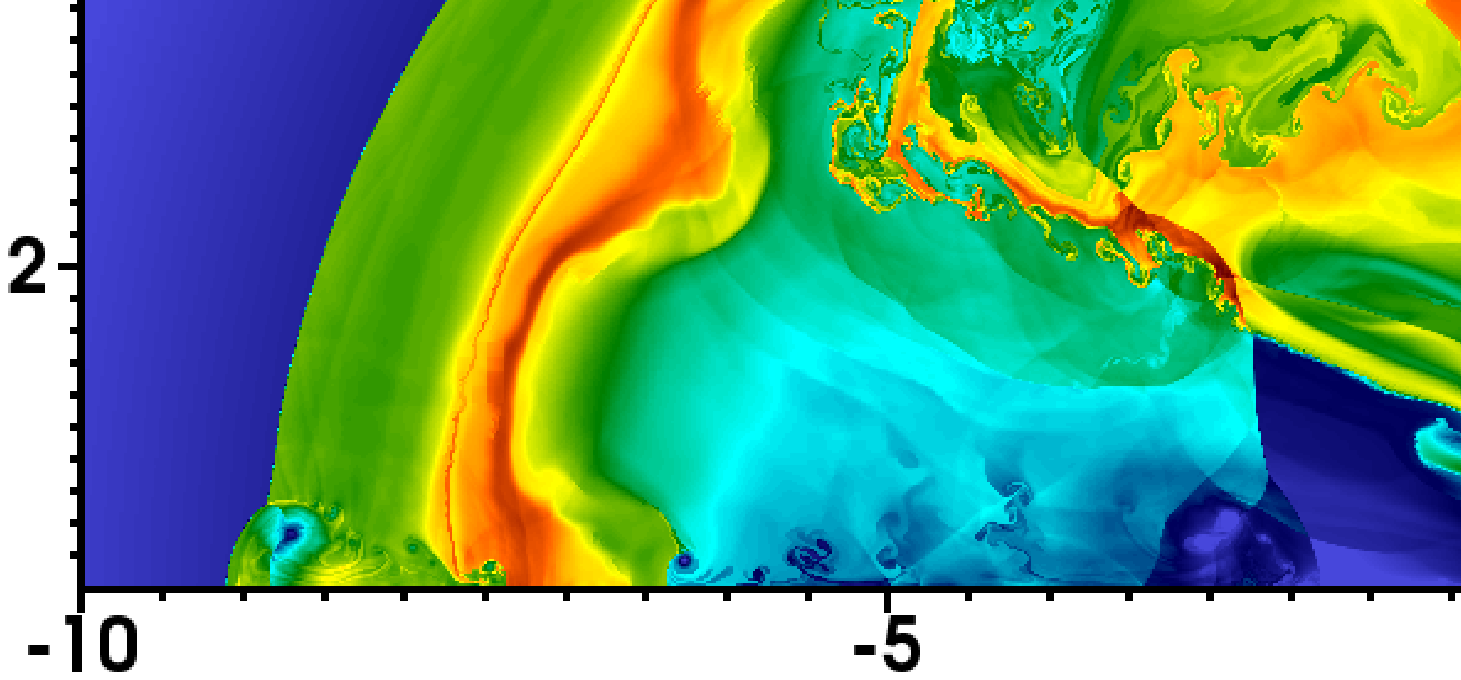}
    \includegraphics[bb=75 571 1154 1099,width=0.45\textwidth,clip=true]
    {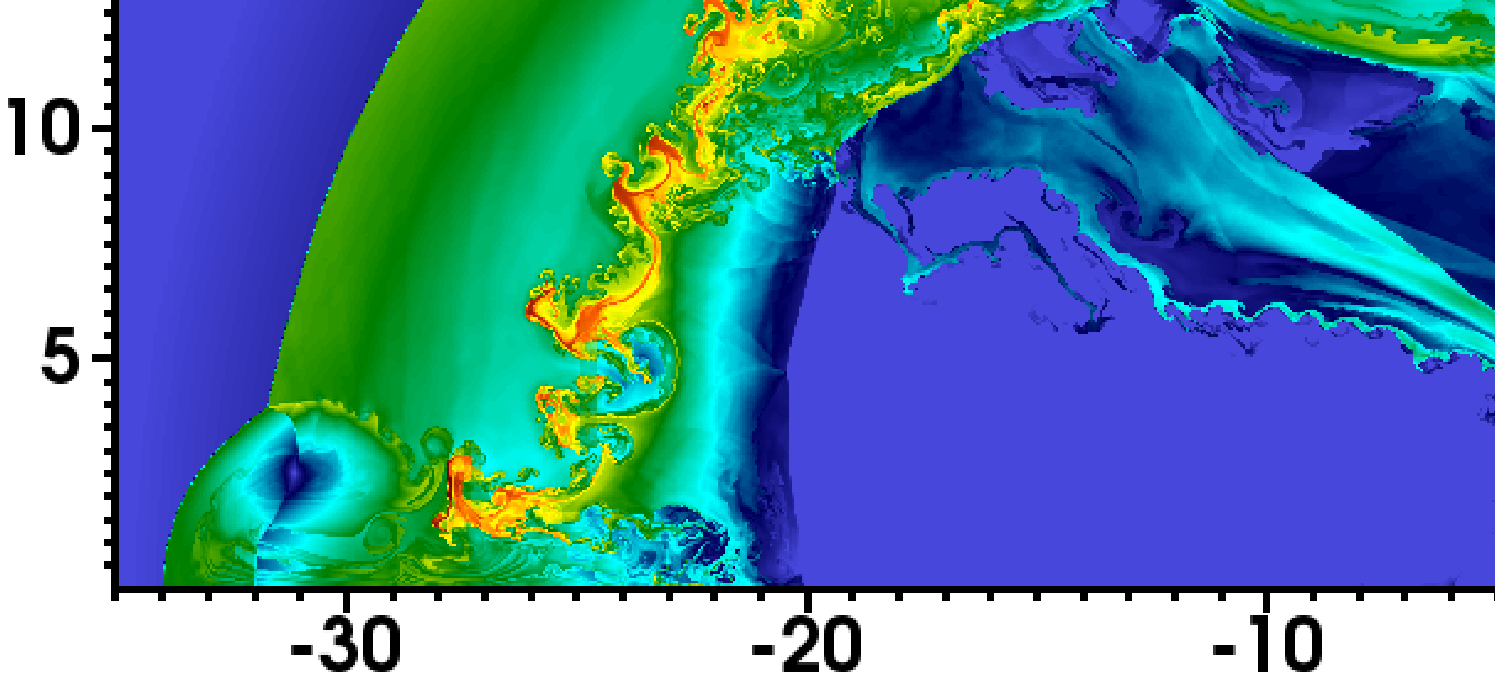}
    \includegraphics[bb=75 571 1154 1099,width=0.45\textwidth,clip=true]
    {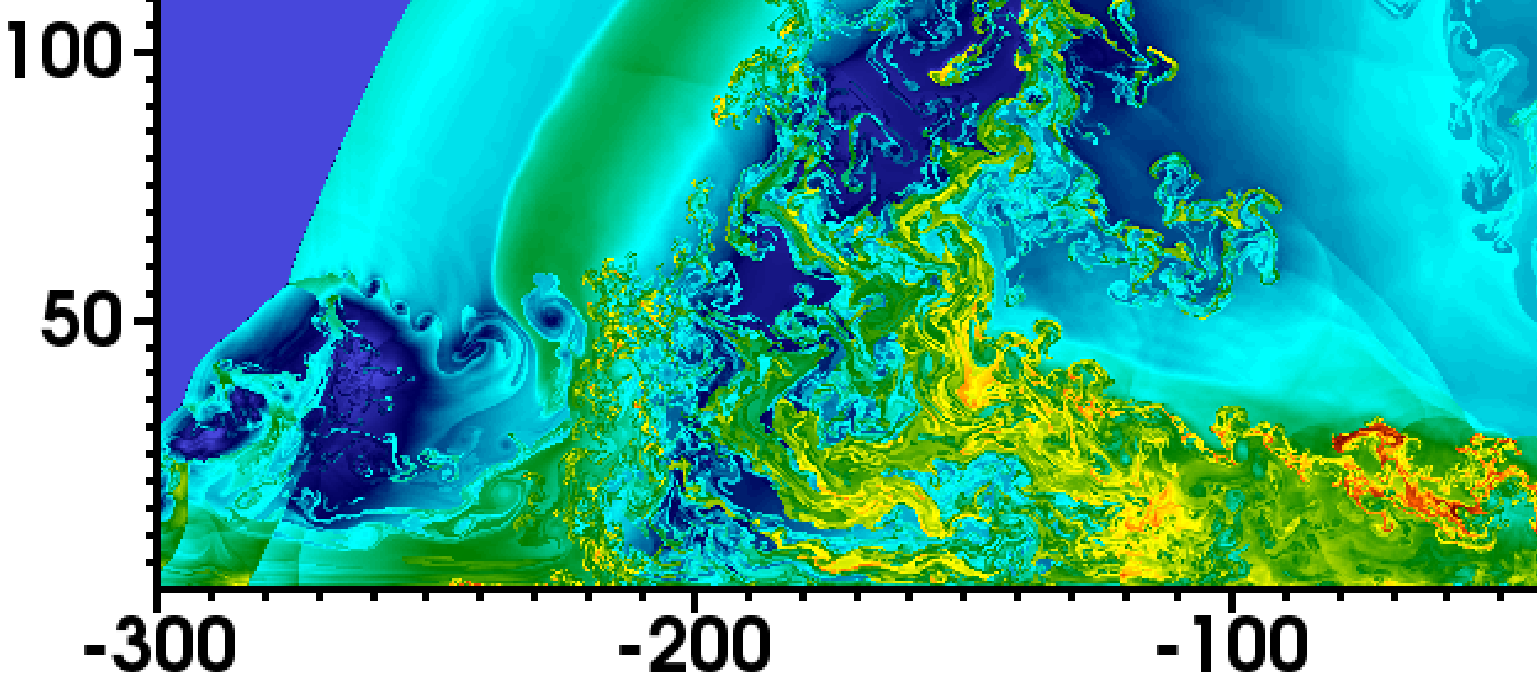}
    \caption{Density distribution in log scale in the $15$~km
    resolution model. (top) $t=4$~s; (middle) $t=20$~s; (bottom)
    $t=300$~s. Note the presence of two post-SASI bubbles, wrapped in
    dense metal-rich material, and several triple points (kinks) along
    the forward shock front at early times; the dense shell features
    well-developed RTI decorations at $t=20$~s.}
    \label{f:snm_r15be_early_cl}
  \end{center}
\end{figure}
%%\end{comment}
%
%
%
shows the density distribution at selected times during the first 5
minutes after core bounce. At $t=4$~s (top panel in
Fig.~\ref{f:snm_r15be_early_cl}), the central region around
$(z,R)=(0,0)$~cm is occupied by the sink hole representing the nascent
neutron star. The material located right outside the hole and in the
equatorial plane is falling back onto the neutron star. The supernova
shock is highly deformed with two low-density bubbles occupying most
of the core region, a clear reminiscence of the low-mode SASI
instability in the b23a model (see also Fig.~3(c) of Paper~II). The
shock front shows several kinks (triple points), the most pronounced
are visible at $(z,R)=(-2\times10^9,5.8\times10^9)$,
$(7.5\times10^9,2.6\times 10^9)$, and $(2\times10^8,6.6\times
10^9)$~cm. As time progresses, those triple points move along the
shock surface, and are sites of strong vorticity deposition at the
material interfaces, significantly enhancing mixing.

%20s
At $t=20$~s (Fig.~\ref{f:snm_r15be_early_cl}, middle panel) the
supernova shock has already traversed through the (C+O)/He interface
and Rayleigh-Taylor instabilities are now fully developed in the dense
shell of material bounded by the supernova shock and the resultant
reverse shock. The kinks in the supernova shock front are clearly
visible although their positions are now slightly different. We notice
clear signs of the Kelvin-Helmholtz instability developing at the
contact surface associated with the leftmost triple point near
$(-1.1\times10^{10},2.15\times 10^{10})$. Two large streams of matter
form a Y-shaped feature along the equator and feed the neutron
star. The edges of both streams show pronounced KHI decorations.

%300s
During the following few hundred seconds, the supernova shock becomes
progressively less aspherical while the dense central region undergoes
dramatic morphological changes. By $t=300$~s
(Fig.~\ref{f:snm_r15be_early_cl}, bottom panel), the supernova shock
has just passed the He/H interface and sweeps through the hydrogen
envelope. The slowing shock causes the formation of a ``helium wall''
behind it. Two pronounced ``scars'' are visible at the wall at
$(-1.4\times 10^{11},1.7\times 10^{11})$ and at $(6.5\times
10^{11},1.85\times 10^{11})$. These flow features are imprints left by
the passing triple points. They will play a crucial role in the deep
mixing of the outer layers into the central regions with large amounts
of hydrogen already penetrating into the helium shell.

At this time we also see that the Rayleigh-Taylor instabilities have
grown substantially. In two regions, highly decorated RTI ``fingers''
have managed to pass through the layers which are in the process of
forming the reverse shock at the inner surface of
the helium wall. The third finger-like feature is located closer to
the equatorial plane and extends half way through the low density core
region. The Y-shaped accretion feature, clearly noticeably at earlier
times, can now be barely recognized as it becomes more extended
laterally and suffers from flow instabilities.

The dense material that penetrates into the layers of the forming
helium wall is of particular interest, as it is composed of
fast-moving, metal-rich clumps. Heavy elements are also
dominant in the jet-like polar outflows visible near $R=0$ and
penetrating all the way up to the supernova shock. The origin of
these features is discussed in
Sect.~\ref{sect:lateral_flow_dynamics}.

\begin{figure*}[ht!]
  \begin{center}
%   \includegraphics[bb=75 571 1154 1099,width=0.45\textwidth,clip=true]
%   {figures/color/snm_r15be_1500s_cl.ps}%
%   \includegraphics[bb=75 571 1154 1099,width=0.45\textwidth,clip=true]
%   {figures/color/snm_r15be_5000s_cl.ps}
%   \includegraphics[bb=75 571 1154 1099,width=0.45\textwidth,clip=true]
%   {figures/color/snm_r15be_10000s_cl.ps}%
%   \includegraphics[bb=75 571 1154 1099,width=0.45\textwidth,clip=true]
%   {figures/color/snm_r15be_1d_cl.ps}
    \includegraphics[bb=75 571 1154 1099,width=0.45\textwidth,clip=true]
    {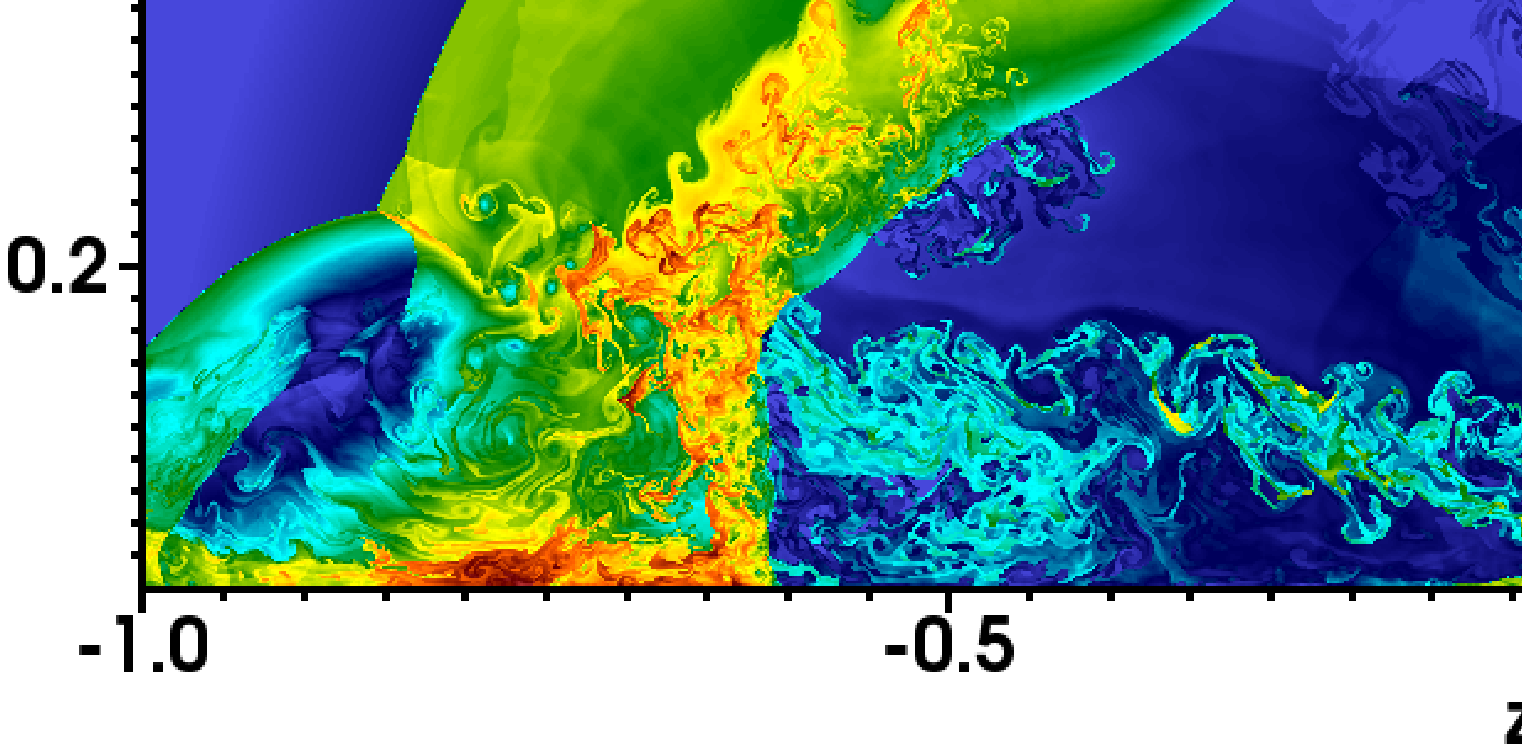}%
    \includegraphics[bb=75 571 1154 1099,width=0.45\textwidth,clip=true]
    {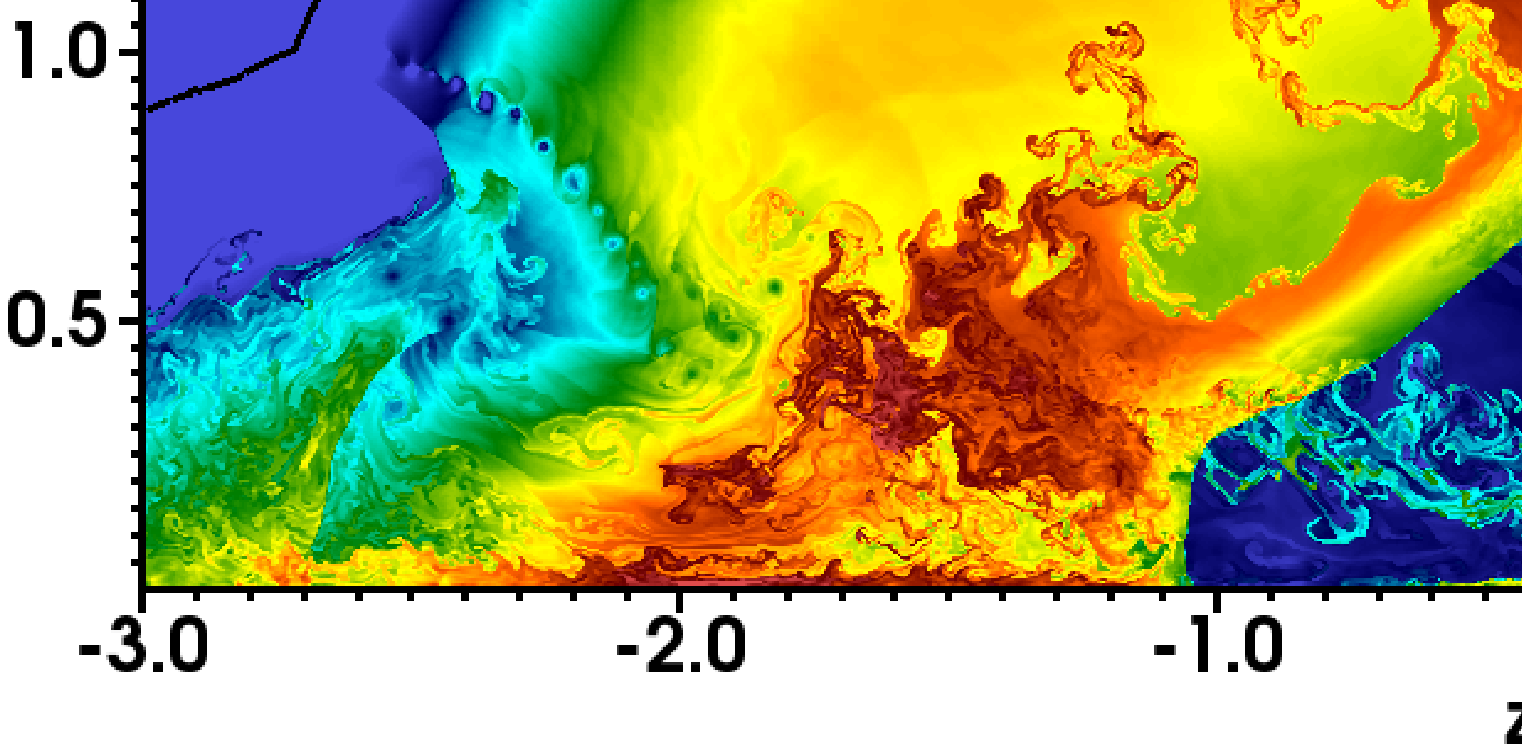}
    \includegraphics[bb=75 571 1154 1099,width=0.45\textwidth,clip=true]
    {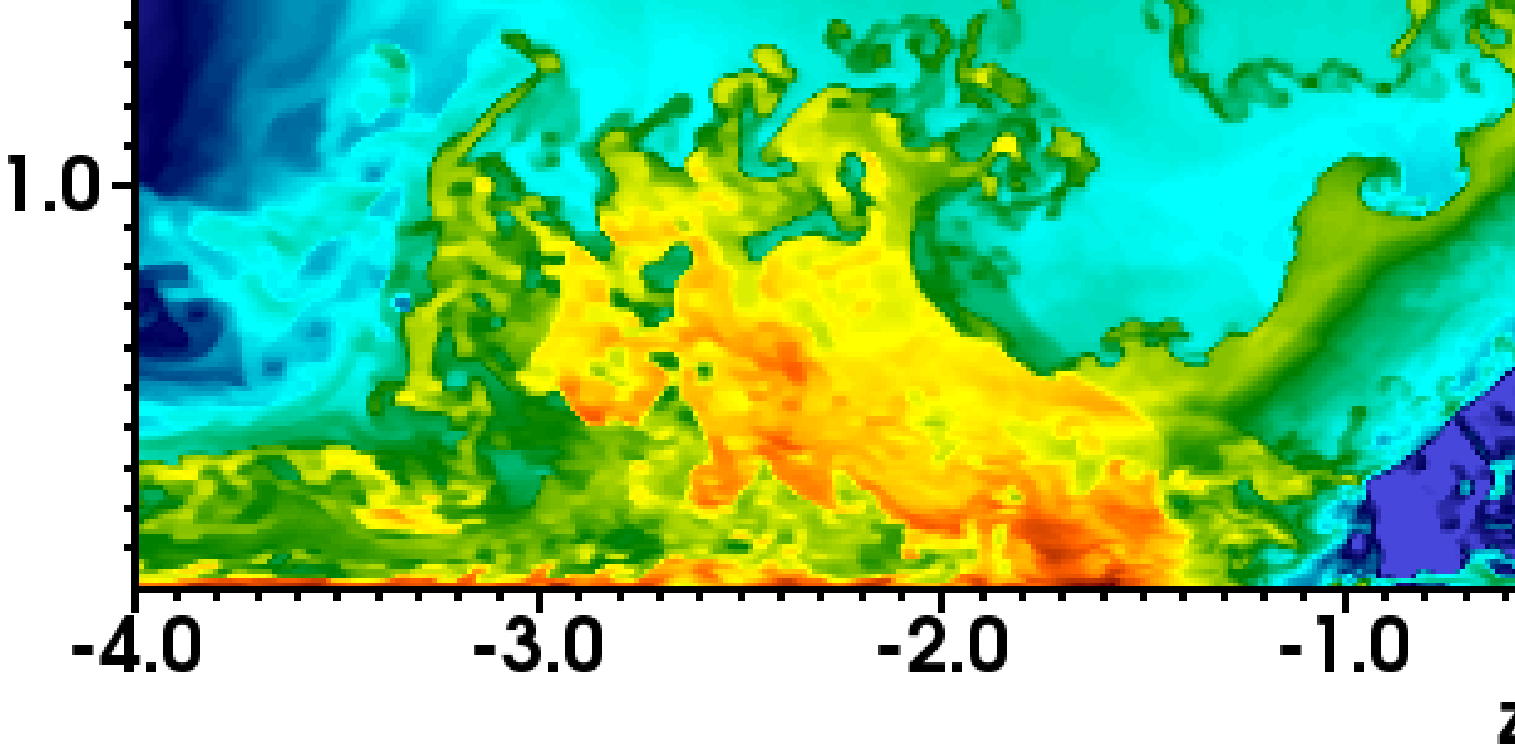}%
    \includegraphics[bb=75 571 1154 1099,width=0.45\textwidth,clip=true]
    {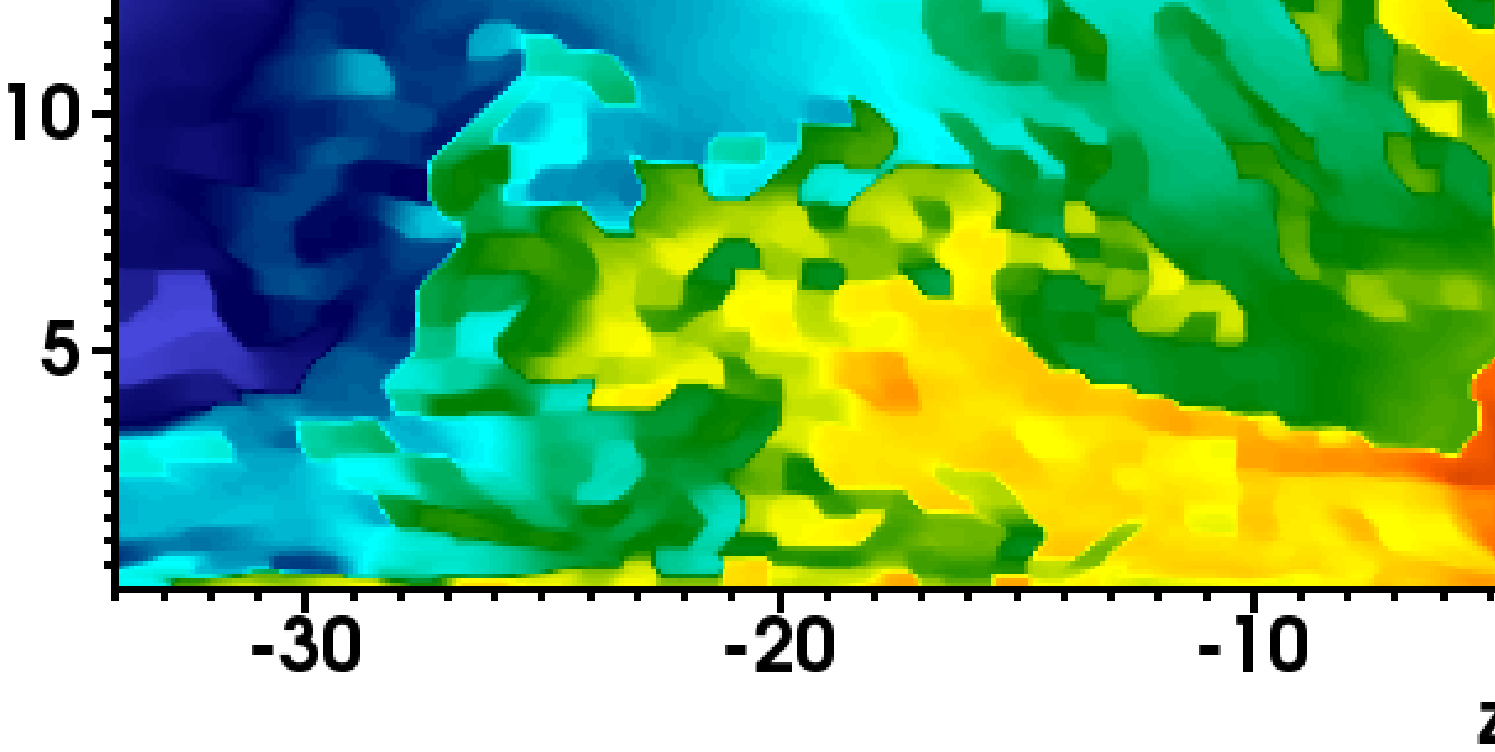}
    \caption{Density distribution in log scale in the $15$~km resolution
    model. (top left) $t=1500$~s; (top right) $t=5000$~s; (bottom
    left) $t=10,000$~s; (bottom right) $t=1$ day. The polar outflows
    are clearly visible at $t=5000$~s, affecting model
    morphology for angles $< 25^\circ$ and $< 20^\circ$ from the
    symmetry axis for negative and positive vertical distances,
    respectively. The supernova shock position at that time is shown
    with a black solid line. Note that the core morphology is
    qualitatively similar for $t\ge 5000$~s.}
  \label{f:1500To1day}
  \end{center}
\end{figure*}

The supernova structure between $t=1500$~s and $t=1$~d is shown in
Fig.~\ref{f:1500To1day}.
%
%
%
%\begin{comment}
%\end{comment}
%
%
%
It can be described as a composite of the supernova shock and its
post-shock region, the mixing layer populated with a network of dense,
metal-rich structures and inward-penetrating hydrogen-rich bubbles,
and the central, low density ejecta being swept by the reverse shock
formed at the base of the helium wall. At $t=1500$~s (top-left panel
in Fig.~\ref{f:1500To1day}), the reverse shock is fully formed and
is about to overrun the middle RTI feature. Hydrogen bubbles are now
more clearly visible, especially the one in the upper left section of
the ejecta. Most of the fast moving metal-rich material is trapped
inside the helium wall although some material
$(z,R)=(3.2\times 10^{11},5.8\times 10^{11})$ appears successful in
penetrating all the way into the hydrogen envelope.

By $t=5000$~s (top-right panel in Fig.~\ref{f:1500To1day}) the
supernova shock is accelerating in its motion through the outer parts
of the stellar envelope. It breaks through the stellar surface at
$t\approx 6000$~s. Substantially more metal-rich clumps are now moving
through the hydrogen envelope and both hydrogen bubbles are fully
developed. There are signs of a mild RTI developing at the outer
surface of the helium wall, best visible close to the equator. The
reverse shock is sweeping through the low density core overrunning the
network of the RTI-fragmented, metal-rich shell in the process. The
central part of the ejecta undergoes relatively little changes
later. Most metal-rich features visible at $t=5000$~s can be clearly
identified also at $t=10,000$~s, and $t=1$ day (bottom-left and
bottom-right panel in Fig.~\ref{f:1500To1day}, respectively). The only
major change is due to the passage of the reverse shock through the
central region of the ejecta. This event does not affect the overall
structure of the core, however.

\begin{figure}[ht!]
  \begin{center}
    \includegraphics[bb=75 571 1154 1099,width=0.45\textwidth,clip=true]
    {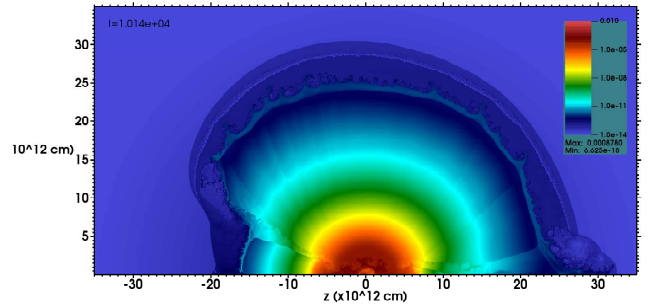}
    \caption{Density distribution in log scale in the $15$~km
      resolution model at $t=10,000$~s including the supernova
      shock. Note the Rayleigh-Taylor instability developing at the
      interface between the shocked stellar ejecta and the shocked
      wind at a distance $r\approx 2.4\times 10^{13}$~cm from the
      explosion center. Note that at this intermediate time, the model
      morphology at large radii is clearly affected by the
      polar outflows for angles $< 35^\circ$ and $< 15^\circ$ from the
      symmetry axis for negative and positive vertical distances,
      respectively.}
    \label{f:snm10000s+fsh}
  \end{center}
\end{figure}

Although at these late times the ejecta morphology appears
well-established, it is the outer regions of the young supernova
remnant where a significant RTI finally comes into play.  This is due
to the final deceleration of the supernova shock inside the stellar
wind: the less dense shocked wind material is now moving slower than
the shocked and denser outer stellar layers. The interface separating
the shocked stellar envelope from the shocked stellar wind is
RTI-unstable \cite[see][and references
therein]{chevalier+92,nymark+06} and numerous RTI fingers are formed
in the process (Fig.~\ref{f:snm10000s+fsh}).
%
%
%
%\begin{comment}
%\end{comment}
%
%
%

\begin{figure}[ht!]
  \begin{center}
    \includegraphics[bb=84 224 573 644,width=0.45\textwidth,clip=true]{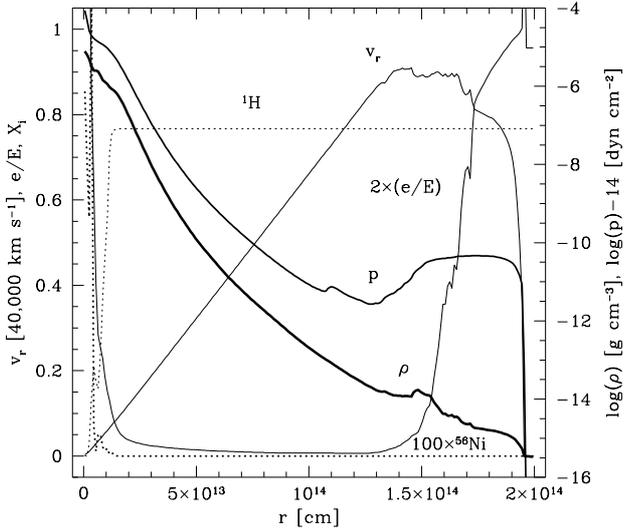}
    \caption{Angle-averaged structure of the 15~km model at $t=12$
      hours. Shown is the density (thick solid), pressure (medium
      solid), radial velocity (thin solid), hydrogen mass fraction
      (thin dotted), nickel mass fraction (thick dotted), and ratio of
      internal to total energy (thin solid). The RTI-unstable
      interface between the shocked ejecta and the shocked stellar
      wind is located around $r\approx 1.5\times 10^{14}$~cm. The
      metal-rich core extends from the center up to $r\approx
      1.5\times 10^{13}$~cm. Note that additional scaling factors are
      used for the pressure, nickel abundance, and the energy ratio.}
    \label{f:12hours}
  \end{center}
\end{figure}

Figure~\ref{f:12hours}
%
%
%
%12hours
%\begin{comment}
%\end{comment}
%
%
%
shows the angle-averaged radial structure of the young supernova
remnant at $t=12$ hours. The outermost RTI-mixing region is clearly
visible in the density, velocity and pressure profiles for
$r=[1.3\times 10^{14},1.7\times 10^{14}]$~cm. Note that at this time
the metal-rich core occupies only $\approx 10$\% of the innermost
region ($r < 2\times10^{13}$~cm). The ejecta are now approaching
homologous expansion as indicated by the low ratio of the internal to
the total energy. The only exception is the freshly shocked
RTI-unstable outermost region. The relatively large internal energy
content of the core ($\approx 10$\% at $r=1\times 10^{13}$cm, and a few
percent at the core's edge) is due to the recent passage of the
reverse shock; the core continues to expand ballistically and will
quickly cool adiabatically.

\begin{figure}[ht!]
  \begin{center}
      \includegraphics[bb=84 224 514 644,width=0.45\textwidth,clip=true]{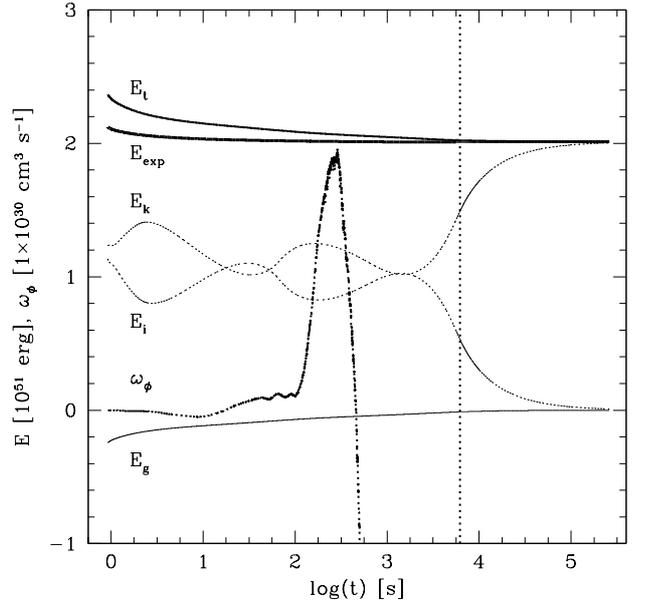}
      \caption{Temporal evolution of integrated quantities in
      the two-dimensional $15$~km resolution model. (thick solid)
      explosion energy; (medium solid) total energy (sum of the
      internal and kinetic energy); (thin solid) gravitational energy;
      (thin dotted) internal energy; (medium dotted) kinetic energy;
      (thick dotted) integrated $\phi$-component of the vorticity,
      $\omega_\phi$. The data are shown up to the moment when the
      supernova shock leaves the computational domain at $t\approx
      2.6\times10^5$~s.}
      \label{f:totals_15km}
  \end{center}
\end{figure}

The low overall internal energy content is confirmed in
Fig.~\ref{f:totals_15km}
%
%
%
%\begin{comment}
%\end{comment}
%
%
%
in which we show the evolution of the model's energies and the
integrated vorticity up to the moment when the shock leaves the
computational domain. The internal and kinetic energy vary together in
sync, reflecting the variations in the shock speed (cf.\
Fig.~\ref{f:shockSpeed}). They are both $\approx 10^{51}$ ergs as
long as the shock remains inside the star. The internal energy rapidly
decreases after the shock breakout through the stellar photosphere,
and amounts to $\approx 1.5$\% of the total energy at $t\approx 23$
hours, when the shock leaves the grid. It is interesting to note the
large variations of the vorticity during the shock breakout. These
variations are due to the rapid increase in the shock speed and slight
global asphericity of the shock front. The vorticity evolution prior
to shock breakout is consistent with that of the models presented in Paper~II.
A direct comparison of our
Fig.~\ref{f:totals_15km} with Fig.~5 of Paper~II is
not possible, though, as the latter displays the evolution of the
vorticity in inner layers of the ejecta (i.e.\ up to the He/H
interface) for the early phase \emph{after} shock breakout. In passing
we note that a realistic model of shock breakout must include
radiation effects \cite[][and references therein]{calzavara+04}, which
are not considered in this work. We expect, however, that these
effects will not affect mixing of the metal-rich core nor the
structure of the young supernova remnant emerging within the first few
days after explosion.

\begin{figure}[ht!]
  \begin{center}
    \includegraphics[bb=75 571 1154 1099,width=0.45\textwidth,clip=true]{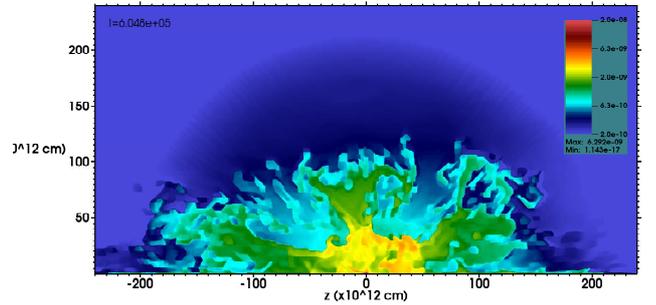}
    \caption{Density distribution in log scale in the $15$~km resolution
      model at $t=7$ days.}
    \label{f:density2d7days}
  \end{center}
\end{figure}

The density distribution in the model at the final time, $t=7$ days,
is shown in Fig.~\ref{f:density2d7days}.
%
%
%
%\begin{comment}
%\end{comment}
%
%
%
Except for the apparent density increase in the central region due to
the reverse shock, the overall structure of the metal-rich part of the
ejecta is quite similar to that obtained a few hours after the core
bounce, and the major high-velocity features can be easily identified
already at $t=5000$~s (see Fig.~\ref{f:1500To1day}). We also note that
the origins of some ejecta features, in particular those of the
hydrogen-rich bubbles, can be tracked all the way back to the first
few minutes after the core bounce. 

The most conspicuous feature visible in Fig.~\ref{f:density2d7days},
though, is the pronounced anisotropy of the ejecta. These have
approximately the form a prolate ellipsoid with a major to minor axis
ratio around 1.6. This was already pointed out in Paper~II based on
simulations that were evolved to a time of 20\,000 seconds after core
bounce. Figure~\ref{f:density2d7days} demonstrates that this result
holds even for the evolution into the homologous phase. In fact,
Fig.~\ref{f:density2d7days} might be regarded as giving an accurate
picture of the ejecta morphology even beyond that phase since the
ensuing expansion is expected to be self-similar, i.e. the morphology
will remain frozen in the flow.

%-------------------------------------------------------------------------
%-------------------------------------------------------------------------
\section{Discussion}
\label{sect:discussion}
%-------------------------------------------------------------------------
%-------------------------------------------------------------------------

%------------------------
\subsection{The SASI-induced lateral expansion: Flow dynamics}
\label{sect:lateral_flow_dynamics}
%------------------------

\begin{figure*}[ht!]
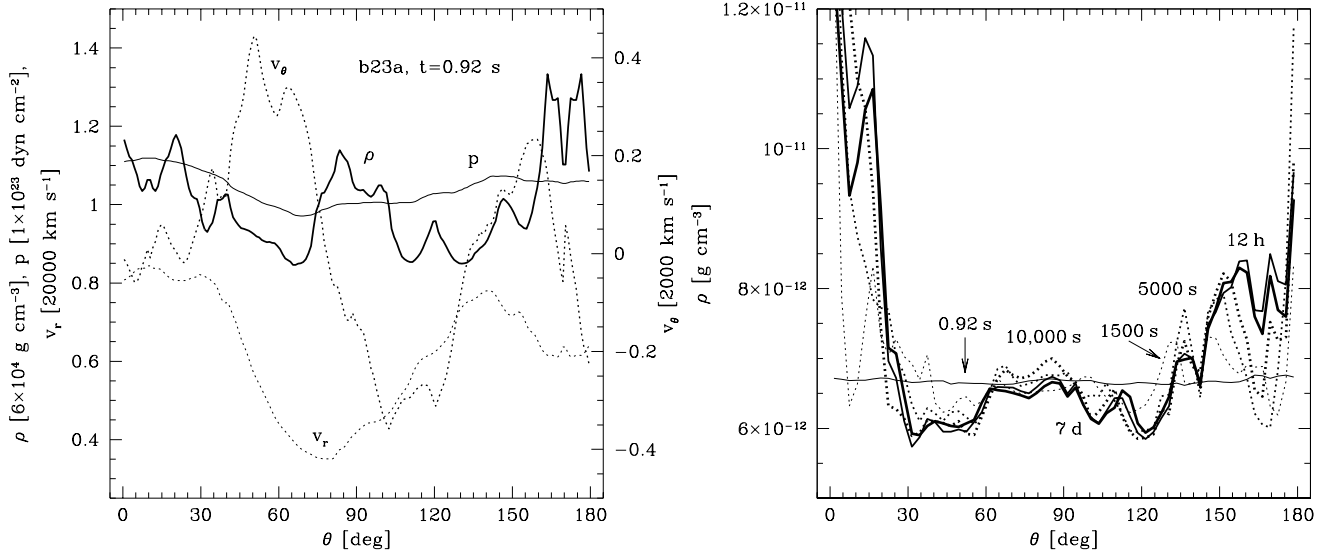

  \begin{center}
    \includegraphics[bb=61 224 570 645,height=0.3\textheight,clip=true]{13431fg17.epsi}%
    \includegraphics[bb=43 224 513 645,height=0.3\textheight,clip=true]{13431fg18.epsi}
    \caption{Angular structure of the $15$~km resolution model. (left
      panel) radial averages of density (thick solid), pressure (thin
      solid), radial velocity (thin dotted), and lateral
      $\theta$-velocity component (thick dotted) at the initial time,
      $t=0.92$~s; (right panel) radial average of density at $0.92$~s
      (thin solid), $1500$~s (thin dotted), $5000$~s (medium dotted),
      $10,000$~s (thick dotted), $12$ hours (medium solid), and $7$
      days (thick solid). Radial averages in the right panel include
      only data for a distance $\leq 1.007\times 10^{15}$~cm from the
      origin. Note the gradual evacuation of the equatorial regions of
      the model and the pile-up of material near the poles
      (right panel), and the strongly anisotropic lateral
      velocity distribution in the explosion model (left panel, thick
      dotted line) characteristic of a low-mode SASI convective
      pattern.}
    \label{f:ExplDensAngular}
  \end{center}
\end{figure*}

A very important effect that we find in the present simulations, whose
consequences have not been recognized previously (although they were
present in the simulations of Paper~II), is the presence of a large
lateral velocity gradient in our initial SASI explosion model (cf. the
left panel of Fig.~\ref{f:ExplDensAngular}). Due to this gradient the
model shows a strong lateral expansion, away from the equator and
toward the poles which leads to an accumulation of material in the polar
regions, within minutes to hours after core bounce (see the right
panel of Fig.~\ref{f:ExplDensAngular}).

\begin{figure}[ht!]
  \begin{center}
    \includegraphics[bb=84 224 569 645,width=0.45\textwidth,clip=true]
    {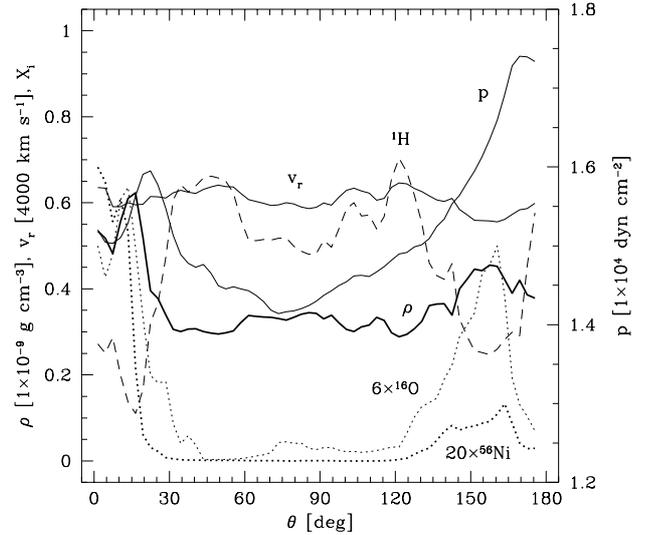}
    \caption{Angular structure of the $15$~km resolution model at the
      final time. Radial averages are shown for the density (thick solid),
      pressure (medium solid), radial velocity (thin solid), and mass
      fraction of hydrogen, oxygen, and nickel (dashed, thin dotted,
      and thick dotted line, respectively). Radial averages include
      only data for distances $\leq 2.4\times 10^{14}$~cm from the
      origin.}
    \label{f:7days_15km_2d_angular}
  \end{center}
\end{figure}

Figure~\ref{f:7days_15km_2d_angular} -- which depicts the angular
distribution of several radially-averaged quantities in the young
supernova remnant at $t=7$ days -- demonstrates that this anisotropy
is preserved until the end of our simulations. Note that the density
(shown with a thick solid line in Fig.~\ref{f:7days_15km_2d_angular}
and in the left panel of Fig.~\ref{f:ExplDensAngular}) is highest
inside the polar regions. Note also that the \emph{initial} angular
distribution of the density is essentially uniform (shown with a thin
solid line in the right panel of Fig.~\ref{f:ExplDensAngular}).

The global lateral expansion contributes decisively to shape the
ejecta into the final form visible in Fig.~\ref{f:density2d7days}. It
is also essential for the formation of the bulges of material that are
visible at the poles of the exploding star as early as $t=300$~s into
the evolution (bottom panel in Fig.~\ref{f:snm_r15be_early_cl}). These
axial flows are the result of a complex interaction between the strong
SASI lateral expansion, and the fact that reflecting boundaries are
used at the symmetry axis which restrict the degrees of freedom of the
flow. Being continuously fed by the lateral motion, and encountering
the impenetrable boundary at the axis, the path of least resistance
for the fluid is to expand along the poles.

In the past, these axial flows have been regarded to be mainly the
result of discretization errors near the axial coordinate singularity,
and conical sections near the poles were used to exclude these regions
from further analysis. However, the pivotal role of the SASI-induced
lateral expansion, indicates that this interpretation does not account
for the true contribution of the various involved effects. Without the
strong continuous feeding by the global lateral flow, the axial
outflows cannot become that strong. This is indicated by the fact that
the SASI-dominated explosions which are considered in both the present
paper and Paper~II show very pronounced axial flows, while the models
presented in Paper~I, where the SASI was absent, showed much weaker
axial features. Consistent with this observation is the fact that in
the limit of a completely vanishing lateral flow, as it is for
instance the case in 2D simulations of Sedov blast waves on spherical
grids, such axial outflows are \emph{not} present.

\begin{figure*}[ht!]
  \begin{center}
%     \includegraphics[bb=97 224 530 640,
%     height=0.3\textheight,clip=true]{figures/mni56_15km_e_2d_angular.epsi}
%     \includegraphics[bb=85 224 585 640,
%     height=0.3\textheight,clip=true]{figures/7days_15km_e_2d_vni56.epsi}
      \includegraphics[bb=97 224 530 640,
      height=0.3\textheight,clip=true]{13431fg20.epsi}
      \includegraphics[bb=85 224 585 640,
      height=0.3\textheight,clip=true]{13431fg21.epsi}
  \end{center}
    \caption{Distribution of nickel in the $15$~km resolution
      model. (left panel) The angular distribution of the nickel mass
      is shown at $0.92$~s (thin dotted), $20$~s (thick dotted),
      $300$~s (thin solid), $5000$~s (medium solid), and $t=7$ days
      (thick solid). {\bf Note the fast advection of nickel to the
      north pole ($\theta \approx 0^{\circ}$).}  (right panel)
      Angular distribution of the nickel mass as a function of
      distance from the origin (right axis) and the radial velocity
      (left axis) {\bf at $t=7$ days}. The nickel mass and average radial velocity are
      calculated using bins with $\Delta r=2.4\times 10^{13}$~cm and
      $\Delta\theta=3^\circ$. The most abundant bins are marked, with
      a nickel mass of at least $1\times 10^{-4}$~\msun\ (solid
      circles) and $1\times 10^{-5}$~\msun\ (open circles). The
      average radial velocity for bins located at the same radial
      distance is shown with dotted lines. Note again the large
      amounts of high velocity nickel located near the symmetry axis
      at $\theta=0^\circ$ and the less massive concentration at
      $\theta\approx 145^\circ$.}
      \label{f:nickel}
\end{figure*}

\subsection{The SASI-induced lateral expansion: Heavy elements redistribution}
\label{sect:nickel_redistribution}

Another very important effect of the SASI-generated lateral expansion
is that it causes a lateral redistribution of material enriched in
heavy elements. Figure~\ref{f:nickel} shows the evolution of the
angular distribution of the nickel mass in our 15~km resolution 2D
model. The initial SASI explosion model contains two nickel-rich
sectors centered around $\theta\approx 45^\circ$ and $\theta\approx
90^\circ$ (thin dotted line in the left panel of
Fig.~\ref{f:nickel}). During the subsequent evolution, the
SASI-generated lateral velocity field of this model
(cf.\ Fig.~\ref{f:ExplDensAngular}) causes a steady drift of the
nickel toward the polar regions.

Of the two nickel concentrations, the material initially located at
$\theta\approx 45^\circ$ is actually the fastest moving nickel-rich
material on our grid, with velocities approaching $4000$~\kms\ (see
the right panel of Fig.~\ref{f:nickel}). This material is advected
over half a quadrant of the computational domain and to the north pole
in less than 300 seconds after core bounce (thin solid line in the
left panel of Fig.~\ref{f:nickel}).

The evolution of the nickel-rich sector initially located near the
equator is somewhat more complex as it eventually distributes nickel
in a broad region between $\theta\approx 135^\circ$ and $\theta\approx
165^\circ$. There is a relatively well-defined nickel-rich ``clump''
(actually a torus-like structure, given that our simulations make use
of axisymmetry) at $\theta\approx145^\circ$ moving with a velocity of
$\approx 2500$~\kms.

In addition to becoming enriched with nickel, the polar regions are
relatively rich in oxygen (thin dotted line in
Fig.~\ref{f:7days_15km_2d_angular}) but devoid of hydrogen (dashed
line in Fig.~\ref{f:7days_15km_2d_angular}). This lateral advection of
heavy elements poses difficulties for the analysis of our 2D
simulations, as it transports the nickel clumps over time into regions
of the computational domain where discretization errors are expected
to be non-negligible and the accuracy of the solution is largely
unknown. There is significant ambiguity in what should and what
shouldn't be accounted for in these regions when computing the maximum
nickel velocities. Given that it is not just numerical errors which
are contributing, but a true physical effect (i.e. the global lateral
expansion) is involved too, it is for instance not at all clear
whether the use of an exclusion cone around the symmetry axis is a
justified and viable approach, or how large that exclusion cone should
be made. This ambiguity is impossible to overcome in axisymmetric
models. It is of a particular consequence since in our highest
resolution 2D model it affects the highest velocity nickel clumps
present in our 2D simulations which are of importance for explaining
the data of SN 1987A.

\begin{figure}[ht!]
  \begin{center}
    \includegraphics[bb=74 345 530 645,width=0.45\textwidth,clip=true]
    {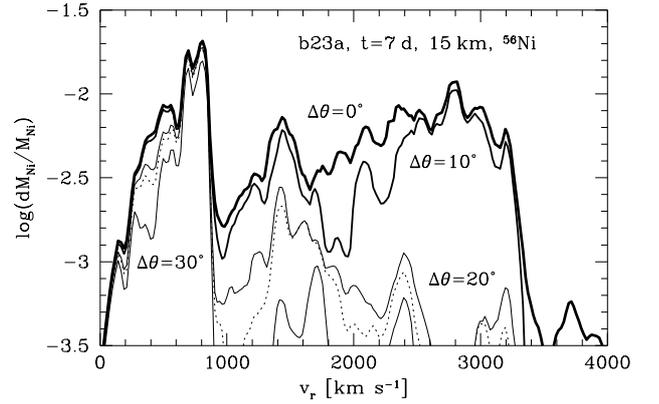}
    \caption{Mass distribution of the nickel in radial velocity space
      in the two-dimensional $15$~km resolution model at $t=7$~days.
      Data for different widths of the polar exclusion cone $\Delta
      \theta$ (solid: $0^\circ$, $10^\circ$, $20^\circ$, and
      $30^\circ$; dashed: $22.5^\circ$)
      are shown.}
    \label{f:nickel_7days}
  \end{center}
\end{figure}

To illustrate this further we show in Fig.~\ref{f:nickel_7days} the
mass distribution of the nickel in velocity space for our 15 km
resolution 2D model at $t=7$~days, as obtained by employing exclusion
cones of different widths. The maximum nickel velocity as determined
without using any exclusion cone at all ($\Delta \theta = 0^\circ$),
is close to 4000~\kms. This value decreases gradually to 3350 \kms,
3250 \kms, and 3200 \kms\ if cones with a width $\Delta \theta$ of
$10^\circ$, $20^\circ$, and $30^\circ$ are used, respectively. The
error in determining this velocity could thus be as high as 25\%,
depending on what one considers to be a true feature of the solution.

%------------------------
\subsection{Extent of the mixing of different elements
\label{ss:mixing}}
%------------------------

\begin{figure*}[ht!]
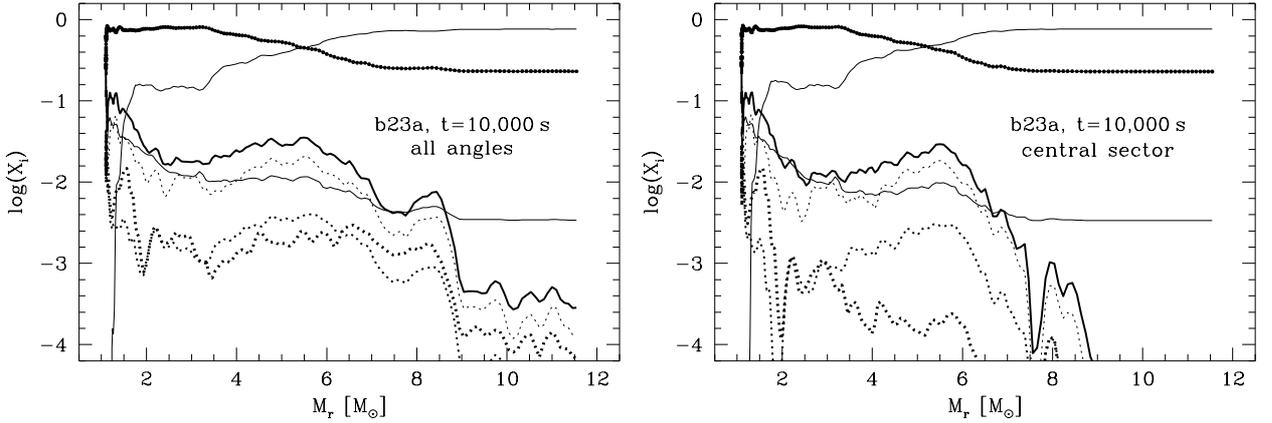

  \begin{center}
    \includegraphics[bb=86 345 514 639,width=0.45\textwidth,clip=true]
    {13431fg23.epsi}
    \includegraphics[bb=86 345 514 639,width=0.45\textwidth,clip=true]
    {13431fg24.epsi}
    \caption{Composition of the two-dimensional $15$~km model at
      $t=10,000$~s. (left panel) entire computational domain; (right
      panel) $135^\circ$ wedge centered at the model equator ($22.5^\circ$
      wedges near the symmetry axis removed). Mass fractions in log
      scale are shown as a function of the radial mass coordinate, $M_r$, for
      hydrogen (thin solid), \nuc{4}{He} (thin with solid circles),
      \nuc{12}{C} (medium solid), \nuc{16}{O} (thick solid),
      \nuc{28}{Si} (thin dotted), \nuc{44}{Ti} (medium dotted), and
      \nuc{56}{Ni} (thick dotted). Note that the equatorial region shows no
      mixing of heavy elements beyond $M_r\approx 9$~\msun.} 
    \label{f:lgxMass}
  \end{center}
\end{figure*}

In evaluating the extent of the mixing of different elements in mass
we again show results considering both the entire computational
domain, and a $135^\circ$ wide wedge centered around the equator
($z=0$~cm) (thereby excluding a cone with width $\Delta \theta =
22.5^\circ$ around the symmetry axis). The latter approach is similar
to what has been done in previous work (though the conical section
used, e.g., in Paper~II had an extent of only $15^\circ$).

Figure~\ref{f:lgxMass} compares the amount of mixing in the $15$~km
model as inferred from the entire domain (left panel) to the mixing in
the region limited to the equatorial wedge (right panel). As we can
see, essentially all material mixed beyond $M_r\approx 9~\msun$\ is
associated with the polar regions of the model. Slightly increasing
the extent of the exclusion cone, i.e.\ making the equatorial wedge
still narrower, does not change the above picture significantly.

Comparing the species distribution as a function of mass at
$t=10,000$~s (right panel in Fig.~\ref{f:lgxMass}) with Fig.~6 of
Paper~II, we observe that at this intermediate time the extent of
mixing in our present high-resolution two-dimensional simulation
matches the amount reported in Paper~II very well. For example the
average position of the He/H interface given in Paper~II is $\approx
5~\msun$, essentially identical to that obtained in our 15\,km
model. In both studies H is mixed down to $1.3~\msun$ while
\nuc{56}{Ni} is mixed out to $\approx 9~\msun$. The agreement is thus
excellent. 

\begin{figure}[ht!]
  \begin{center}
    \includegraphics[bb=86 346 513 640,width=0.45\textwidth,clip=true]
    {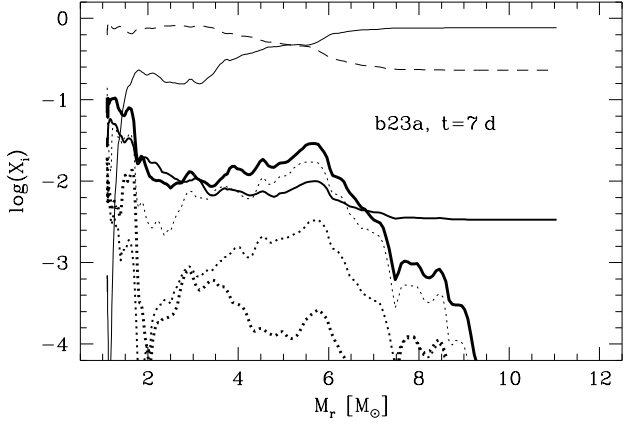}
    \caption{Composition in the equatorial section (polar regions
      excluded) of the two-dimensional $15$~km resolution model at
      $t=7$ days.  Mass fractions in log scale are shown as a function
      of the radial mass coordinate for hydrogen (thin solid),
      \nuc{4}{He} (dashed), \nuc{12}{C} (medium solid), \nuc{16}{O}
      (thick solid), \nuc{28}{Si} (thin dotted), \nuc{44}{Ti} (medium
      dotted), and \nuc{56}{Ni} (thick dotted).}
    \label{f:lgx_mass_7d_15km_wedge}
  \end{center}
\end{figure}

By comparing the species distribution at $t=10,000$~s (right panel in
Fig.~\ref{f:lgxMass}) to that at $t=7$ days
(Fig.~\ref{f:lgx_mass_7d_15km_wedge}), we can, moreover, see that most
of the mixing takes place within the first 2 or 3 hours of the
supernova explosion. This observation is consistent with the ratio of
the internal to total energy decreasing to $\approx 0.15$ by
$t=10,000$~s (see Fig.~\ref{f:EintotAngular} and the discussion in
Sect.~\ref{ss:homology}). At that time, lateral pressure gradients are
very small, and the fluid elements move nearly ballistically.

\begin{figure}[ht!]
  \begin{center}
    \includegraphics[bb=74 345 530 645,width=0.45\textwidth,clip=true]
    {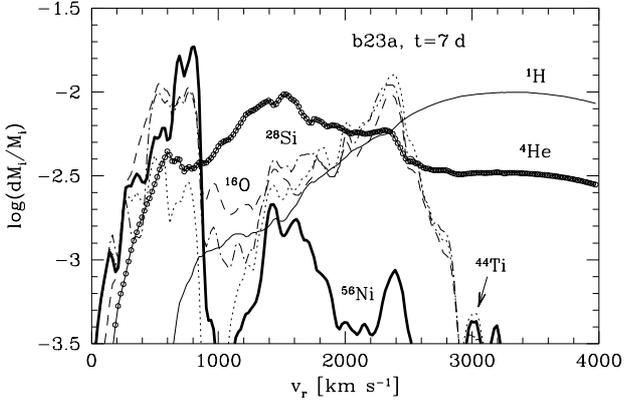}
    \caption{Mass distributions in radial velocity space in the
      two-dimensional $15$~km resolution model at $t=7$~days. The
      distribution of mass is shown for hydrogen (thin solid),
      \nuc{4}{He} (thin solid with open circles), \nuc{16}{O}
      (dashed), \nuc{28}{Si} (dash-dotted), \nuc{44}{Ti} (dotted), and
      \nuc{56}{Ni} (thick solid). Only data for the equatorial section
      of the model are shown (polar regions are excluded).}
    \label{f:lgxvr7days}
  \end{center}
\end{figure}

As it has been discussed extensively in Paper~II, both the inward
mixing of hydrogen and the outward mixing of nickel are crucial for an
understanding of SN 1987A's data. The inward mixing of hydrogen down
to $1.3~\msun$ is required for explaining the broad peak of the light
curve and the very low minimum hydrogen velocities of this supernova,
which according to \cite{Kozma_Fransson98} were $ \lesssim700$ \kms.
Figure~\ref{f:lgxvr7days}, which displays mass distributions in
velocity space for the different nuclear species at the final
simulation time, demonstrates that this is excellently reproduced by
our models. Again an exclusion cone around the symmetry axis with
$\Delta \theta = 22.5^\circ$ has been used for producing this
figure. The outward mixing of nickel to a mass coordinate of $9~\msun$
is tantamount to the presence of nickel-rich material on the grid with
velocities $\gtrsim 3200$ \kms. We will return to this point in
Sect.~\ref{ss:convergence}, where we will discuss the dependence of
the nickel velocity distribution on the numerical resolution.

%------------------------
\subsection{Evolution towards homology}
\label{ss:homology}
%------------------------

\begin{figure}[ht!]
  \begin{center}
    \includegraphics[bb=84 224 516 639,width=0.45\textwidth,clip=true]
    {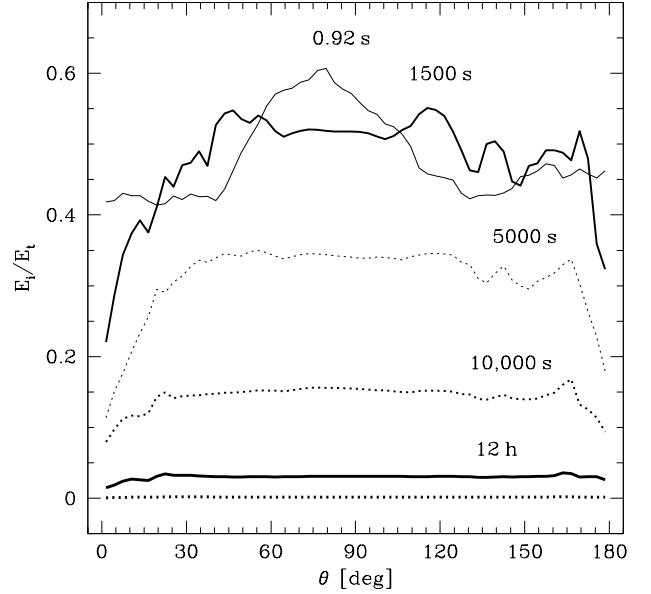}
    \caption{Evolution of the internal to total (internal plus
      kinetic) energy ratio in the $15$~km resolution model. Radial
      averages of the energy ratio are shown at $0.92$~s (thin solid),
      $1500$~s (medium solid), $5000$~s (thin dotted), $10,000$~s
      (medium dotted), $12$ hours (thick solid), and $7$ days (thick
      dotted). Radial averages include only data for distances $\leq
      2.517\times 10^{14}$~cm from the origin. The apparent excess
      heat content of the explosion model (thin solid line) near the
      equator is due to low fluid velocities in a funnel separating
      two rolls (bubbles) which were created by the SASI in the initial
      phases of the explosion.}
    \label{f:EintotAngular}
  \end{center}
\end{figure}

The evolution of the system towards homology is characterized by the
ratio of internal to total energy. As we discussed in
Sect.~\ref{ss:2d}, the internal energy of the exploding supernova
varies during the early stages of shock propagation through the
envelope. It begins to decrease steadily for $t\ge 1500$~s (see
Fig.~\ref{f:totals_15km}), when the supernova shock starts its final
acceleration phase prior to breaking out through the stellar surface.
At this time the internal energy amounts to about half of the total
energy of the supernova and exhibits substantial variations in angle
(medium solid line in Fig.~\ref{f:EintotAngular}).

\begin{figure*}[ht!]
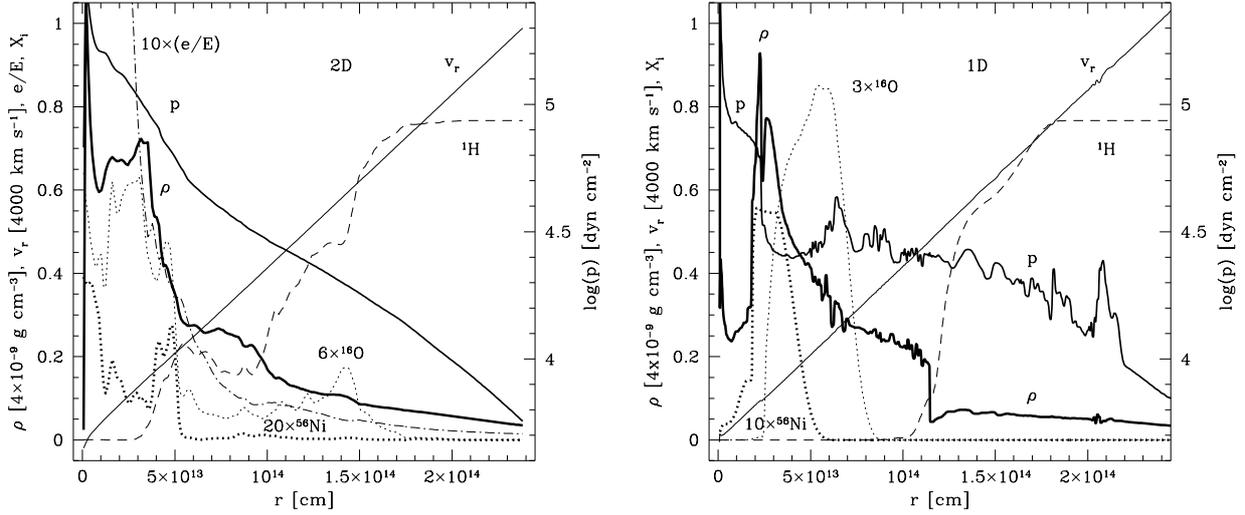

%%%  \begin{center}
  \centering
    \includegraphics[bb=85 224 586 640,width=0.45\textwidth,clip=true]
    {13431fg28.epsi}
    \includegraphics[bb=85 224 586 640,width=0.45\textwidth,clip=true]
    {13431fg29.epsi}
%%%  \end{center}
  \caption{Structure of the exploding supernova in the $15$~km resolution
    models at the final time of $t=7$\, days. 
    (left panel) Two-dimensional model. (right
    panel) One-dimensional model. In both panels we show the distributions of
    density (thick solid), pressure (medium solid), radial velocity
    (thin), hydrogen mass fraction (dashed), \nuc{16}{O} mass fraction
    (thin dotted), and \nuc{56}{Ni} mass fraction (thick dotted line). In
    addition, the ratio of internal to total energy in the
    two-dimensional model is shown in the left panel (dash-dotted).}
  \label{f:densityMore7days}
\end{figure*}

These are largely smoothed out by $t=5000$~s (thin dotted line in
Fig.~\ref{f:EintotAngular}). The average internal energy content drops
to $\approx 15$\% by $t=10,000$~s, and well below 5\% by $t=12$
hours. It is unclear whether this remaining energy is sufficient to
modify the expansion velocities of our 15\,km model. Using a polar
exclusion cone of $22.5^\circ$ width around the poles, we observe a
relatively modest increase of the maximum nickel velocity on the grid
by about 12\%, from $\approx 2850$~\kms\ at $t=10,000$~s (not shown)
to $\approx 3200$~\kms\ at $t=7$ days
(Fig.~\ref{f:lgxvr7days}). However, and as explained in
Sect.~\ref{sect:nickel_redistribution}, at these late times the
nickel-rich material in question is located so close to the north
pole, that the determination of the maximum nickel velocity is
affected by the ambiguity in choosing the width of the exclusion cone
(see also Sect.~\ref{ss:convergence}). With the morphology of the flow
remaining essentially unchanged for $t > 5000$~s, it is, however, safe
to state that, for the employed blue supergiant progenitor model, the
ejecta become homologous during the first day of the evolution.

%------------------------
\subsection{Comparison to one-dimensional models}
%------------------------

As a final illustration of the importance of the multidimensional
effects in our models we contrast angle-averaged radial profiles
obtained from our 15 km resolution two-dimensional model (left panel
in Fig.~\ref{f:densityMore7days}) with plots of our corresponding
spherically symmetric (1D) model with the same resolution and at the
final time (right panel in Fig.~\ref{f:densityMore7days}).

Both models appear to produce explosions of comparable energies as
indicated by the radial velocity profiles. Both models are also quite
similar for radii $\ge 2\times 10^{14}$~cm, i.e.\ in regions
unaffected by mixing in multidimensions. However, at smaller radii, we
observe an extensive mixing of nuclear species. For example,
substantial amounts of oxygen are present at $r\approx
1.5\times10^{14}$~cm, and hydrogen is mixed down to $r\approx
3\times10^{13}$~cm in the multidimensional model. The density jump
associated with the H/He interface, which is clearly visible at
$r\approx 1.15\times10^{14}$~cm in 1-D, is completely smeared in
multidimensions. The only source of mixing in 1-D is numerical
diffusion, while hydrodynamic instabilities -- which are partly caused
by the initial shock nonuniformity -- contribute to mixing in
multidimensions.

Note the presence of regions of opposite density and pressure
gradients, as required by the Rayleigh-Taylor instability, in the
spherically symmetric model at $r\approx 6\times 10^{13}$~cm and
$r\approx 1.1\times 10^{14}$~cm. The former region is associated with
the outer edge of the nickel-rich region, while the latter is
connected to the H/He interface; both regions experienced vigorous
mixing in multidimensions, as expected.

\subsection{Numerical convergence in multidimensions}
\label{ss:convergence}

Our final concern is mesh convergence (see Sect.~\ref{ss:1d} for a
discussion of the one-dimensional case). The quantity which is of
largest importance for comparison to observations and which shows the
most sensitive dependence on the mesh resolution of our 2D models is
the nickel distribution. In contrast, the distribution of
e.g. hydrogen is very similar in all of our 2D models. We will
therefore focus on the nickel distribution in what follows.

It is important to note that for the present problem proper
convergence analysis is made very difficult by several effects. First,
no analytic solution for the problem is known, making it impossible to
determine the actual solution error. Second, the problem is highly
non-linear. Thus, even if one would obtain an estimate of the
instantaneous solution error by Richardson extrapolation, the
contribution and interaction of higher order error terms, which are
not captured by this technique, must be considered highly likely. An
estimate of the instantaneous solution error is, moreover, not
relevant as we are primarily interested in the accuracy of the
solution at late times, when the error has accumulated. And third, in
the context of the present axisymmetric simulations, significant
difficulties and ambiguities for the analysis of the simulations are
encountered near the poles.

\begin{figure*}[ht!]
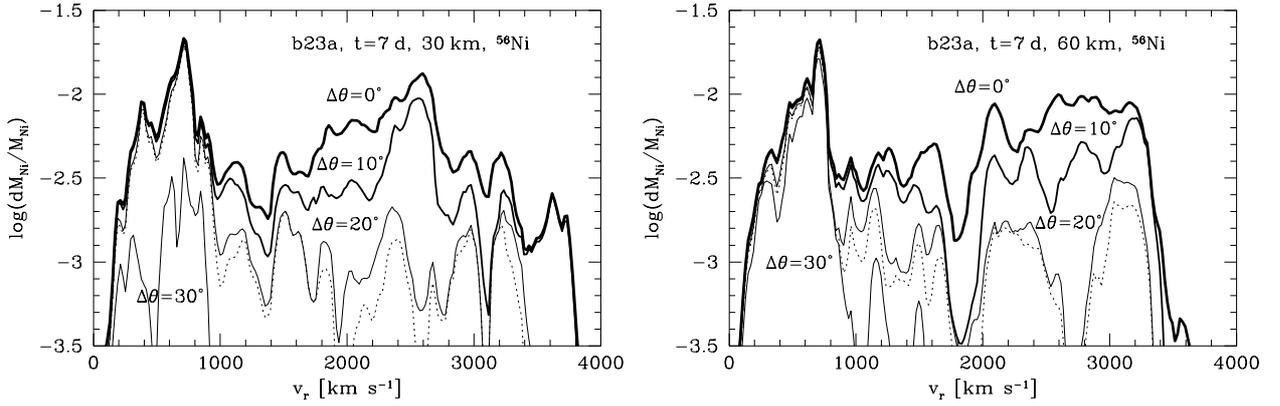

  \begin{center}
  \includegraphics[bb=74 345 530 645,width=0.45\textwidth,clip=true]
    {13431fg30.epsi}
  \includegraphics[bb=74 345 530 645,width=0.45\textwidth,clip=true]
    {13431fg31.epsi}
    \caption{Same as Fig.~\ref{f:nickel_7days} but for our two-dimensional
             models with 30~km resolution (left panel), and 60~km resolution
             (right panel).}
    \label{f:nickel_7days_30_60km}
  \end{center}
\end{figure*}

In Fig.~\ref{f:nickel_7days} we have already demonstrated that the
high-velocity tail of the nickel distribution of our 15 km resolution
model at $t=7$~days depends sensitively on the width of the exclusion
cone that is used around the symmetry
axis. Figure~\ref{f:nickel_7days_30_60km} shows that the same is true
in our lower-resolved 30~km and 60~km models. Particular noteworthy is
that the maximum nickel velocity as obtained from these figures
depends on both the width of the exclusion cone, $\Delta \theta$, and
on the numerical resolution of the simulation.

\begin{table}
\begin{center}
\caption{Dependence of the maximum nickel velocities (in \kms) 
         in our two-dimensional models with different spatial resolution
         on the width of the polar exclusion cone $\Delta \theta$.}
\label{tab:nickel_vel}
\begin{tabular}{ccccc}
\hline
           & \multicolumn{4}{c}{\mbox{$\Delta \theta$}} \\
Resolution & $0^\circ$ & $10^\circ$ & $20^\circ$ & $30^\circ$ \\
\hline
15 km  & 3950 & 3350 & 3250 & 3200 \\ 
30 km  & 3800 & 3800 & 3450 & 3400 \\
60 km  & 3600 & 3425 & 3350 & 3350 \\
\hline
\end{tabular}
\end{center}
\end{table}

Table~\ref{tab:nickel_vel} summarizes this dependence for
convenience. It demonstrates that for no value of $\Delta \theta$,
convergence in the maximum nickel velocities is observed with
increasing resolution. In fact, the obtained maximum nickel velocities
are not even monotonic if an exclusion cone with a finite width is
used. Only for the case where no exclusion cone at all is employed
($\Delta \theta = 0^\circ$) do we observe monotonically increasing
maximum nickel velocities with increasing resolution, which, however,
do not show a clear sign of convergence.

From Table~\ref{tab:nickel_vel} we can only infer a lower limit for
the maximum nickel velocities at $t=7$~days of $\gtrsim 3200$ \kms. It
is also noteworthy that for a 240~km resolution model, \cite{guzman09}
reported a qualitatively different nickel distribution, with no nickel
moving faster than $\approx 2100$~\kms. In his study, significant
amounts of nickel moving with velocities $>3000$~\kms\ are observed
for the first time at a resolution of 120~km, which is consistent with
what we find here. The maximum velocity of Fe-group elements reported
in Paper~II, where an exclusion cone of $\Delta \theta = 15^\circ$ was
used, is $\approx 3300$~\kms (see Fig.~8 of that work). This is in
good agreement with the above results, although mesh remapping, and a
different mesh geometry were used in this earlier work, and the
evolution was followed to only 20,000 seconds after core bounce.

From Figs.~\ref{f:nickel_7days} and \ref{f:nickel_7days_30_60km} it is
furthermore obvious that also the \emph{amount} of nickel in the
high-velocity tail shows no signs of convergence. Even between the
better resolved models it can differ by up to a factor of three, when
no exclusion cone is used to analyze the simulations. Using exclusion
cones of finite width, the differences become even larger. In
contrast, the mass of heavy elements moving at $1000-2400$~\kms\ is
quite similar in the models presented here: the mass of nickel
expanding with $\approx 2000$~\kms\ converges to within a factor of
2-3 (the same holds in case of the 120~km model obtained by
\citealt{guzman09}). In addition, this result is less sensitive
to the width of the exclusion cone around the symmetry axis.

It should be noted that in addition to the variations introduced by
the numerical resolution, and the ambiguity of analyzing the
simulations near the poles, there are two further uncertainties in the
models that concern the nickel velocities and the initial nickel
distribution. First, our axisymmetric simulations tend to overestimate
the drag coefficient of true three-dimensional clumps
\citep{kane+00,hammer+09}. And second, the abundance ratios of different
iron-group nuclei in neutrino-driven supernova models depend very
sensitively on the neutrino luminosities, and cannot be calculated
accurately with a grey neutrino transport scheme, as e.g. the one that
was employed in calculating model b23a (see Paper~II). In other words,
the $\rm ^{56}Ni$ abundance in model b23a may be lower than in
reality, favoring nuclei like $\rm ^{57}Fe$, $\rm ^{58}Fe$, etc.,
instead. In Paper~II we have tried to compensate for this uncertainty
by considering only the \emph{total} mass fraction of iron group
nuclei in the analysis of velocity distributions. We have proceeded
similarly here -- lumping together different iron group nuclei into
what we called the ``nickel'' abundance -- since the total abundance
of the iron-group is expected to be a more reliable quantity than its
individual abundances, and can thus serve as an upper limit to the
$\rm ^{56}Ni$ abundance. A more detailed investigation of these issues
is required, though, and will be given in future work.

%-------------------------------------------------------------------------
%-------------------------------------------------------------------------
\section{Conclusions}
\label{sect:summary}
%-------------------------------------------------------------------------
%-------------------------------------------------------------------------
We presented the results and detailed analysis of a series of
axisymmetric hydrodynamic simulations of the first week of the
evolution of a non-spherical core-collapse supernova.
Different from our previous work, our computations were performed
in a single domain using cylindrical coordinates. We implemented a
workload-constrained mesh adaption strategy that allowed us to complete
the simulations given limited computational resources, and to avoid the
cumbersome periodic mesh remapping used in past studies
\citep{kifonidis+06,couch+09}. We obtained a series of models at
progressively higher mesh resolution and provided insight into the
numerical convergence of our simulations. 

Our simulations are the first to follow a SASI-dominated explosion
from shortly after its initiation into the homologous expansion and
early SNR phase. Especially during the first several hours of this
evolution, the ejecta are characterized by complex interactions
between Rayleigh-Taylor, Richtmyer-Meshkov, and Kelvin-Helmholtz
instabilities, which produce an extensive mixing and outward
penetration of stellar layers enriched in heavy elements. The global
asphericity of the supernova shock is, moreover, essential for
triggering a deep inward penetration of hydrogen and helium into the
central ejecta regions. For the particular explosion model and
progenitor that we have studied, the ejecta become homologous
approximately one day after the explosion.

A very important result of our present simulations is that the 2D SASI
instability, which acts during the explosion launching phase, not only
determines the structure of our model around that time, but leaves
also a strong, large-scale imprint in the ejecta in the form of a
significant lateral velocity gradient (Fig.~\ref{f:ExplDensAngular}),
which affects the evolution for minutes to hours later (i.e. for several
RT-growth time scales). This means that although the radiative driving
of the explosion essentially ends around 1 second after the core
bounce, it is actually incorrect to consider the late-time evolution
as a phase which is completely detached from the history of the
explosion.

This strong lateral expansion of SASI explosion models sheds new light
onto the formation of the polar outflows which have been observed in
previous axisymmetric simulations of late-time supernova evolution.
It had actually been mentioned in Paper~II that the appearance of
these outflows was likely not due to a single numerical problem, but
rather due to the combination of the restriction of the degrees of
freedom of the flow, the use of reflecting boundaries at the axis
of symmetry, and the presence of a coordinate singularity, and hence
the possible occurrence of non-negligible discretization errors at
this axis.

A crucial finding of our present work is that even this quite
differentiated view for the formation of the polar outflows is
\emph{incomplete}. It neglects the fact that also a \emph{physical}
cause is involved, namely the strong late-time lateral expansion of 2D
SASI explosion models. Our simulations indicate that this is actually
the dominant effect for the formation of such pronounced flows near
the symmetry axis. The strong lateral expansion away from the
equatorial plane and toward the poles has, moreover, two very
important consequences for the observational outcome of the models: it
results in an inevitable advection of high-velocity nickel-rich
material from moderate latitudes toward the poles, and it contributes
to ultimately shape the ejecta into the form of a prolate ellipsoid.

In the present model, the highest velocity nickel on our grid, which
moves with speeds close to 4000 \kms\, and is consistent with the data
of SN 1987A, is advected within only 300 seconds after core bounce
over half a quadrant of the computational domain. In other words,
although this nickel-rich material is initially located far away from
the axis, by the end of the simulation it has ended up close to the
poles. If confirmed in future three-dimensional simulations, this
effect of the SASI might actually explain the asymmetric nickel lines
of SN 1987A.

Yet, since the accuracy of the solution is largely unknown near the
poles, there is significant ambiguity in what should be accounted for
in these regions when computing the maximum nickel velocities of our
2D simulations. Given that a physical effect is involved in
transporting material to the poles, it is for instance not at all
clear whether the use of an exclusion cone at the poles -- which was
customary in earlier simulations -- is indeed justified. It is also
unclear how large such an exclusion cone should be made. This
ambiguity is impossible to overcome in axisymmetric models.

The medium (i.e. 60~km, and 30~km) resolution models and partly also
the high (i.e. 15~km) resolution model that we presented here, yield
results which are in very good agreement with the simulations reported
in Paper~II \citep{kifonidis+06}, although these latter simulations
were obtained with a different code, and a different computational
strategy, and were followed to only 20\,000 seconds after core bounce.
In particular we found that the amount of mixing of both light and
heavy elements, and the maximum nickel velocities in both studies
agree well.

The extent of the mixing of heavy elements in our high-resolution
simulations differs qualitatively from models which are less well
resolved (as e.g.\ the 240~km model obtained by
\citealt{guzman09}). At high resolutions, the mass distribution
displays a pronounced hump with velocities $<1000$~\kms\ and a long
tail extending out to $\gtrsim 3200$~\kms. The hump region shows signs
of convergence while quantitative differences are seen in the tail. At
the end of the simulations, the highest nickel velocities vary between
$3200$~\kms and $\sim 4000$~\kms, while the mass of the fastest moving
nickel varies by at least a factor 2-3 between our best resolved
models. These results are sensitive as to whether the polar regions
are included in the analysis or not. This shows that due to the strong
non-linearity of the problem, and the ambiguity in analyzing
two-dimensional simulations near the axis of symmetry, strict
numerical convergence is difficult to achieve. It also implies that
present three-dimensional simulations, with their much lower
resolution, must be viewed as being far from resolved, and that their
conclusions must be verified by a proper numerical convergence
analysis in three dimensions.

The second important effect of the SASI-induced lateral expansion is
that it contributes significantly to the strong prolate deformation of
the ejecta, as it is observed at the end of our simulations (i.e. at
$t=7$~days). This ejecta deformation may be considered final, because
the expansion has long become homologous by that time. As
\cite{kjaer+10} have pointed out in their recent study, it is moreover
in very good agreement with the ejecta morphology of SN 1987A, making
the assumption of a ``jet-induced'' explosion unnecessary.

Given the importance of the SASI-induced late-time lateral expansion
for both SNR morphology and the distribution of heavy elements,
systematic future studies are required to investigate how it depends
on the dominant SASI mode when the early explosion phase ends. In the
present simulations it is the $l=2$ (quadrupolar) mode of the SASI
which ultimately became dominant, but other modes are clearly possible
in 2D simulations \citep{scheck+06}, while little is known in the
three-dimensional case. Since a thorough survey of SASI mode outcomes
cannot be completed in two dimensions and considering furthermore the
problems that the analysis of 2D simulations poses with regard to the
determination of the maximum nickel velocities, 3D high-resolution
simulations should be performed as soon as corresponding computational
resources will allow for that. Such three-dimensional simulations will
need to employ singularity-free grids that cover the entire sphere, in
order to be free of the aforementioned problems, and hence of optimal
use in both early and long-time supernova modeling.

\begin{acknowledgements}
We thank Paul Drake for helpful discussions and encouragement, and an
anonymous referee for his comments on the manuscript, which have
helped us to improve the clarity of our presentation.  AG was
supported by the Polish Ministry of Science through the grant
92/N-ASTROSIM/2008/0. TP was supported through the DOE grant
DE-FG52-03NA000064.  This research used resources of the National
Energy Research Scientific Computing Center, which is supported by the
Office of Science of the U.S. Department of Energy under Contract No.\
DE-AC02-05CH11231, NASA's Astrophysics Data System, and software in
part developed by the DOE-supported ASC/Alliance Center for
Astrophysical Thermonuclear Flashes at the University of Chicago.
\end{acknowledgements}
%
%
%---------------------------------------------------------------
%---------------------------------------------------------------
%\section*{References}
%---------------------------------------------------------------
%---------------------------------------------------------------
%
%
%
\bibliographystyle{aa}
\bibliography{13431}
\end{document}